\documentclass{article}

\usepackage[preprint, nonatbib]{neurips_2024}

\usepackage[utf8]{inputenc} 
\usepackage[T1]{fontenc}    
\usepackage{hyperref}       
\usepackage{url}            
\usepackage{booktabs}       
\usepackage{amsfonts}       
\usepackage{nicefrac}       
\usepackage{microtype}      
\usepackage{xcolor}         
\usepackage{amsmath}
\usepackage{multirow}
\usepackage[pdftex]{graphicx}
\usepackage{pifont}

\usepackage[backend=biber, maxnames=10]{biblatex}
\usepackage{subcaption}    
\usepackage{array}         
\usepackage{gensymb}       
\usepackage{wrapfig}       

\usepackage[framemethod=TikZ]{mdframed} 
\usepackage[capitalise]{cleveref}      
\usepackage{pifont}                     
\usepackage{enumitem}       

\definecolor{promptcolor}{RGB}{233, 233, 235}
\definecolor{prompttitlecolor}{RGB}{175, 172, 172}
\mdfdefinestyle{prompt}{%
    linecolor =   black,
    linewidth =   1pt,
    backgroundcolor =  promptcolor,
    roundcorner     =   5pt,
    font = \small\ttfamily
}

\newmdenv[%
    roundcorner=5pt, 
    linecolor =   black,
    linewidth =   1pt,
    font = \small\ttfamily,
    subtitlebackgroundcolor=prompttitlecolor, 
    frametitlebackgroundcolor=prompttitlecolor,
    backgroundcolor=promptcolor, 
    frametitle={Generated caption},
    subtitleaboveskip=0.5\baselineskip,
    subtitlebelowskip=0.5\baselineskip,
    ]{captionenv}

\definecolor{citecolor}{RGB}{14, 123, 73}
\definecolor{urlcolor}{RGB}{37, 50, 102}
\definecolor{linkcolor}{RGB}{255, 55, 78}
\hypersetup{
   colorlinks,
   linkcolor={linkcolor},
   citecolor={citecolor},
   urlcolor={urlcolor}
}
\hypersetup{colorlinks}

\newcommand{\cmark}{\textcolor{linkcolor}{\ding{51}}}
\newcommand{\xmark}{\textcolor{citecolor}{\ding{55}}}

\newcommand{\sqm}{m\(^2\)}
\newcommand{\mdeg}{\(\degree\)}

\newbibmacro{string+url}[1]{%
  \iffieldundef{doi}{%
    \iffieldundef{url}{%
      \iffieldundef{isbn}{%
        \iffieldundef{issn}{%
          #1%
        }{%
          \href{http://books.google.com/books?vid=ISSN\thefield{issn}}{#1}%
        }%
      }{%
        \href{http://books.google.com/books?vid=ISBN\thefield{isbn}}{#1}%
      }%
    }{%
      \href{\thefield{url}}{#1}%
    }%
  }{%
    \href{http://dx.doi.org/\thefield{doi}}{#1}%
  }%
}
  
\DeclareFieldFormat{title}{\usebibmacro{string+url}{\mkbibemph{#1}}}
\DeclareFieldFormat[article, inproceedings, book]{title}{\usebibmacro{string+url}{\mkbibquote{#1}}}

\title{Learning Spatially-Aware\\ Language and Audio Embeddings}

\author{%
  Bhavika Devnani\(^{1, }\)\thanks{Work done while at Apple.} \quad Skyler Seto\(^2\) \quad Zakaria Aldeneh\(^2\)\quad Alessandro Toso\(^2\)\\
  \textbf{Elena Menyaylenko}\(^2\) \quad 
  \textbf{Barry-John Theobald}\(^2\) \quad \textbf{Jonathan Sheaffer}\(^2\)\quad \textbf{Miguel Sarabia}\(^2\) \\
   \(^{1}\) Georgia Institute of Technology \qquad \(^{2}\) Apple \\
  \texttt{bdevnani3@gatech.edu, \{sseto, zaldeneh, atoso\}@apple.com}  \\
  \texttt{\{elenam, bjtheobald, sheaffer, miguelsdc\}@apple.com}\\
}

\begin{document}

\maketitle

\begin{abstract}
 
Humans can picture a sound scene given an imprecise natural language description. For example, it is easy to imagine an acoustic environment given a phrase like ``the lion roar came from right behind me!''. For a machine to have the same degree of comprehension,  the machine must know what a lion is (semantic attribute), what the concept of ``behind'' is (spatial attribute) and how these pieces of linguistic information align with the semantic and spatial attributes of the sound (what a roar sounds like when its coming from behind). 
State-of-the-art audio foundation models, such as CLAP~\cite{elizalde2023clap,wu2023large}, which learn to map between audio scenes and natural textual descriptions, are trained on non-spatial audio and text pairs, and hence lack spatial awareness. In contrast, sound event localization and detection models are limited to recognizing sounds from a fixed number of classes, and they localize the source to absolute position (e.g., 0.2m) rather than a position described using natural language (e.g., ``next to me''). To address these gaps, we present ELSA (Embeddings for Language and Spatial Audio), a spatially aware-audio and text embedding model trained using multimodal contrastive learning. ELSA supports non-spatial audio, spatial audio, and open vocabulary text captions describing both the spatial and semantic components of sound. To train ELSA: (a) we spatially augment  the audio and captions of three open-source audio datasets totaling 4,738 hours and 890,038 samples of audio comprised from 8,972 simulated spatial configurations, and (b) we design an encoder to capture the semantics of non-spatial audio, and the semantics and spatial attributes of spatial audio using contrastive learning. ELSA is a single model that is competitive with state-of-the-art for both semantic retrieval and 3D source localization.  In particular, ELSA achieves \(+2.8\%\) mean audio-to-text and text-to-audio R@1 above the LAION-CLAP~\cite{wu2023large} baseline, and outperforms by \(-11.6\)\mdeg~mean-absolute-error in 3D source localization over the SeldNET~\cite{NERCSLIP} baseline on the TUT Sound Events 2018 benchmark \cite{adavanne2018sound}. Moreover, we show that the representation-space of ELSA is structured, enabling swapping of direction of audio via vector arithmetic of two directional text embeddings.

\end{abstract}

\section{Introduction}

Humans use implicit context when communicating about and comprehending sounds in their environment. For instance, the instruction ``Pull over if you hear a siren from behind you'' is easily understood by most humans.
However, a machine would need to not only recognize the source the sound, i.e., the siren (a semantic cue), but also interpret the spatial reference implied by ``behind'' relative to its own position (a spatial cue). The machine must then translate these linguistic cues into its understanding of spatial audio to accurately identify, locate, and conditionally respond to the sound. This degree of alignment between spatial audio and natural language is understudied in prior work.

Audio foundation models (AFMs), such as LAION-CLAP~\cite{wu2023large}, have been used for multiple downstream applications, such as language guided audio editing~\cite{wang2023audit,jiang2024listen}, language guided audio and music generation~\cite{huang2023makeanaudio,ghosal2023texttoaudio, yang2023diffsound}, audio representations for image and text~\cite{guzhov2021audioclip,yariv2023audiotoken}, setting a precedent for the wide applicability of audio representations aligned with natural language.  However, these models, and similar state-of-the-art AFMs, such as Pengi~\cite{deshmukh2023pengi}, LTU~\cite{gong2023listen}, and SALMONN~\cite{tang2023salmonn}, cannot capture the \emph{spatial attributes} as the models are trained only on single-channel/non-spatial audio. Conversely, models such as SELDNet \cite{adavanne2018sound},
and PILOT \cite{schymura2021pilot} are capable of precise spatial attribute classification and regression, but lack capability to generalize to natural language descriptions of spatial and semantic attributes.

To address these challenges, we introduce ELSA, a multimodal foundation model that learns a joint representation space for the spatial attributes and semantics of audio aligned with natural language descriptions of the audio scene. Learning a joint contrastive representation model enables and improves several tasks including: retrieval, multimodal QA, captioning, and generation \cite{wu2023large}.  Prior work has shown the benefits of learning well-aligned encoders for multimodal tasks within the vision domain \cite{merullo2022linearly,shen2021much}.  In contrast, mapping between audio and natural language via a large language model, as in \cite{zheng2024bat}, can significantly improve language reasoning tasks, especially for zero-shot generalization of pre-trained models, however yields worse performance in classification and QA tasks when fine-tuned in the language domain \cite{wang2022language}. ELSA enables similar spatially-aware downstream applications, for instance, one can expand traditional language-guided audio editing to manipulate spatial elements using natural language commands like: ``Remove the sound of the plane flying above'' or ``move the sound of the dog barking from left to right''. In this paper we focus, for the first time, on devising and analyzing multimodal task-agnostic representations that capture both the semantics and the spatial attributes of audio aligned with natural language.

Noting a lack of paired spatial audio and language data that can enable training spatially aware audio-language models at scale, 
to train ELSA, we synthesize a spatial audio corpus consisting of 890,038 samples that span a variety of acoustic room properties, such as size and reverberation, from audio clips of the AudioSet~\cite{gemmeke2017audio} and Freesound~\cite{Freesound2013} corpora.  We also synthesize natural language spatial audio captions to match the spatial audio using a large-language model (LLM) to rephrase the initial captions.  We demonstrate that ELSA captures spatial attributes and semantics of audio by identifying a set of tasks on which a standard AFM, such as LAION-CLAP, fails.  Moreover, we show that ELSA achieves better zero-shot classification of spatial attributes than models trained only for that task.  Finally, we show that ELSA maintains the ability to represent non-spatial audio by demonstrating performance competitive with existing state-of-the-art for a number of tasks.

Our key contributions are:

\begin{itemize}
\item We present and release a new synthetic dataset of 4738.55 hours, with 890,038 samples and corresponding spatial captions across 8,972 simulated rooms with accurate parametric labels for the room properties and sound source locations.  Additionally, we also record a small spatial real-world dataset to verify transfer to the real-world (cf.~\cref{sec:spatial_dataset}).

\item We provide ELSA, a multimodal spatial audio-language model that jointly performs semantic classification (sound detection, retrieval), spatial localization, and direction of arrival. ELSA consists of an audio encoder paired with a text encoder that jointly learns semantic and spatial attributes via contrastive learning (cf.~\cref{sec:pretraining}).

\item We show that ELSA effectively captures spatial attributes and semantics competitive with baselines. ELSA improves by \(-11.6\)\mdeg~mean-absolute-error on 3D source localization, and by \(+2.9\%\)  on text-to-audio and audio-to-text mAP@10 scores. (cf.~\cref{sec:experiments}).

\item Further, we show that the representation-space of ELSA is structured, allowing for transposition of spatial sound direction via addition or subtraction of two spatially descriptive text embeddings. (cf.~\cref{sec:embedding_structure}).

\end{itemize}

\section{Related Work}

We provide an overview of the architecture choices and corresponding datasets for training and evaluation of models that can capture the semantics and spatial attributes of audio, including AFMs.

\paragraph{Audio-language Approaches} Prior works (e.g., CLAP~\cite{elizalde2023clap}, LAION-CLAP~\cite{wu2023large}, MULAN~\cite{huang2022mulan}) have extended the image-text contrastive pre-training approach introduced by CLIP~\cite{radford2021learning} to link audio representations to textual descriptions. These models use two encoders, one for audio and another for text, to project the representations from the two modalities into a common embedding space. Once trained, the models enable zero-shot prediction and retrieval capabilities on unseen sounds and textual descriptions. Despite their utility, the methods do not capture the spatial attributes of the modeled signals, rather they capture only their semantics. Another line of work (e.g., Pengi~\cite{deshmukh2023pengi}, LTU~\cite{gong2023listen}, and SALMONN~\cite{tang2023salmonn}) extends LLMs to enable audio understanding in open-vocabulary settings (e.g., audio captioning and audio question answering). Such models learn audio encoders to provide a prefix token to prompt a frozen pre-trained autoregressive LLM, which is then used to generate unconstrained text. These prior methods do not explicitly model the spatial attributes of the audio. Zheng et al.~\cite{zheng2024bat} introduced BAT, an audio-based LLM that combines binaural spatial sound perception with natural language understanding with an accompanying question-and-answer dataset that enables model training. BAT focuses on enabling LLMs to reason about binaural spatial audio, which depends on the head-related transfer-function.  In contrast our focus is on a task-agnostic and device-agnostic representation of spatial audio aligned with text.

\paragraph{Audio-language Datasets} Learning audio-language models requires access to datasets that link the two modalities. Clotho~\cite{drossos2020clotho} and AudioCaps~\cite{kim2019audiocaps} are popular audio captioning datasets for which the textual descriptions were collected by annotating sound event datasets (e.g., AudioSet~\cite{gemmeke2017audio}, or Freesound~\cite{Freesound2013}) through crowd-sourcing platforms. LAION-Audio-630K~\cite{wu2023large} is a large-scale audio-text dataset collected by downloading audio and relevant textual descriptions from publicly available websites. All three datasets focus on the semantic attributes of the audio signal and do not have labels for the spatial attributes. SPATIALSOUNDQA~\cite{zheng2024bat} is a dataset that consists of simulated binaural audio samples and question-answer pairs, which was used to train BAT~\cite{zheng2024bat}. The audio samples were sourced from AudioSet~\cite{gemmeke2017audio}, and the question-answer pairs were paraphrased using GPT-4. In contrast with the our environmental description captions, the text in SPATIALSOUNDQA is geared towards question-and-answer tasks. In addition, the dataset employs a binaural representation of spatial audio, rendering the data incompatible with ELSA. STARSS23~\cite{starss23} is a dataset of real-world multi-channel audio annotated with semantic labels for overlapping sound sources, and the equivalent annotations for the spatial attributes.  However, a limitation is that the dataset lacks natural language descriptions of the sound scenes, which are required for aligning the spatial attributes with language descriptions.


\section{Paired Spatial Audio and Text Datasets}
\label{sec:spatial_dataset}

\begin{figure}[t]
    \centering
     \begin{subfigure}[t]{0.32\textwidth}
         \centering
         \includegraphics[width=\textwidth]{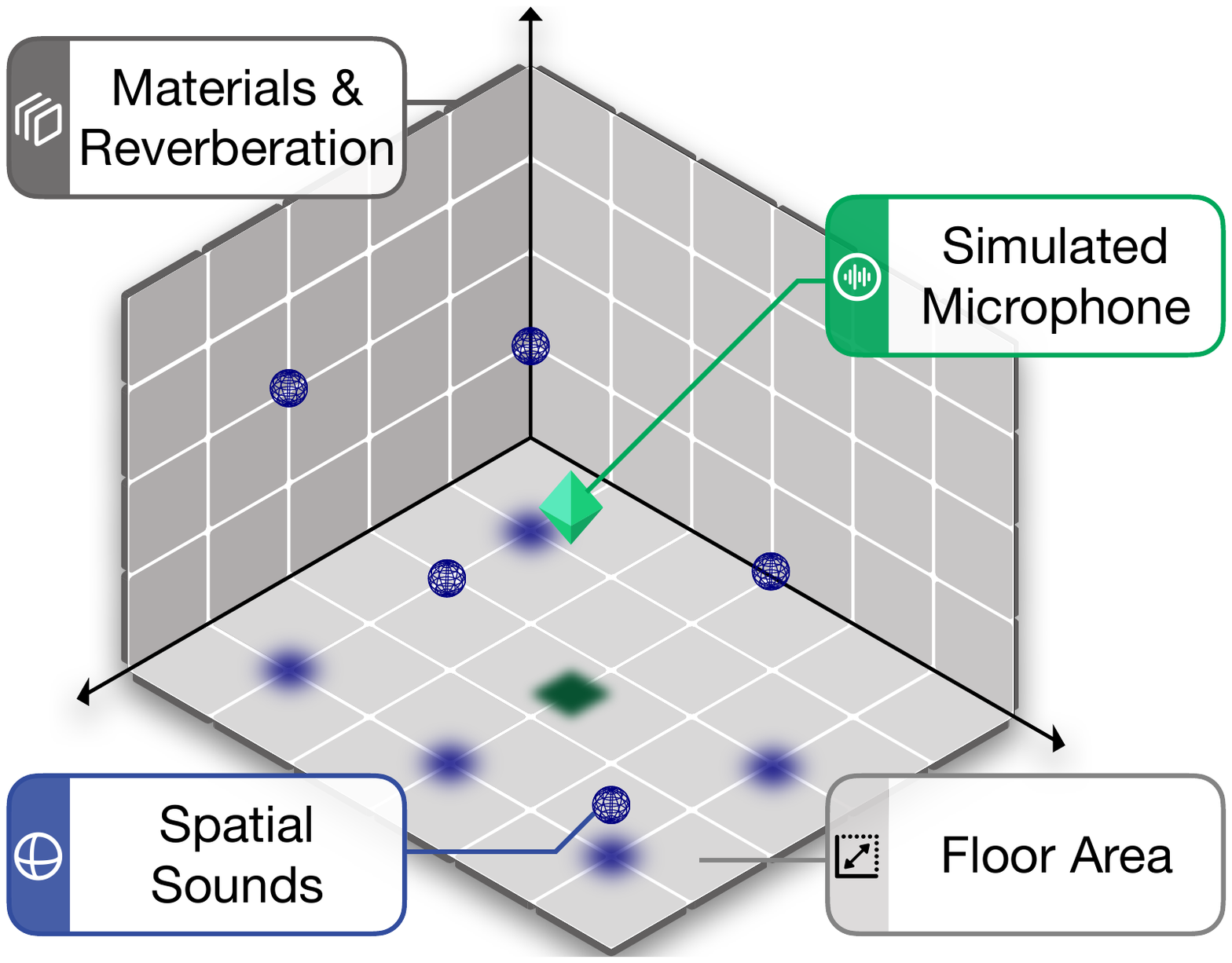}
         \caption{Our spatial audio pipeline uses simulated rooms with different dimensions, materials, and reverberation, and with sources located at different spatial locations.}
         \label{fig:spatial_audio_pipeline}
     \end{subfigure}
     \hfill
     \begin{subfigure}[t]{0.32\textwidth}
         \centering
         \includegraphics[width=\textwidth]{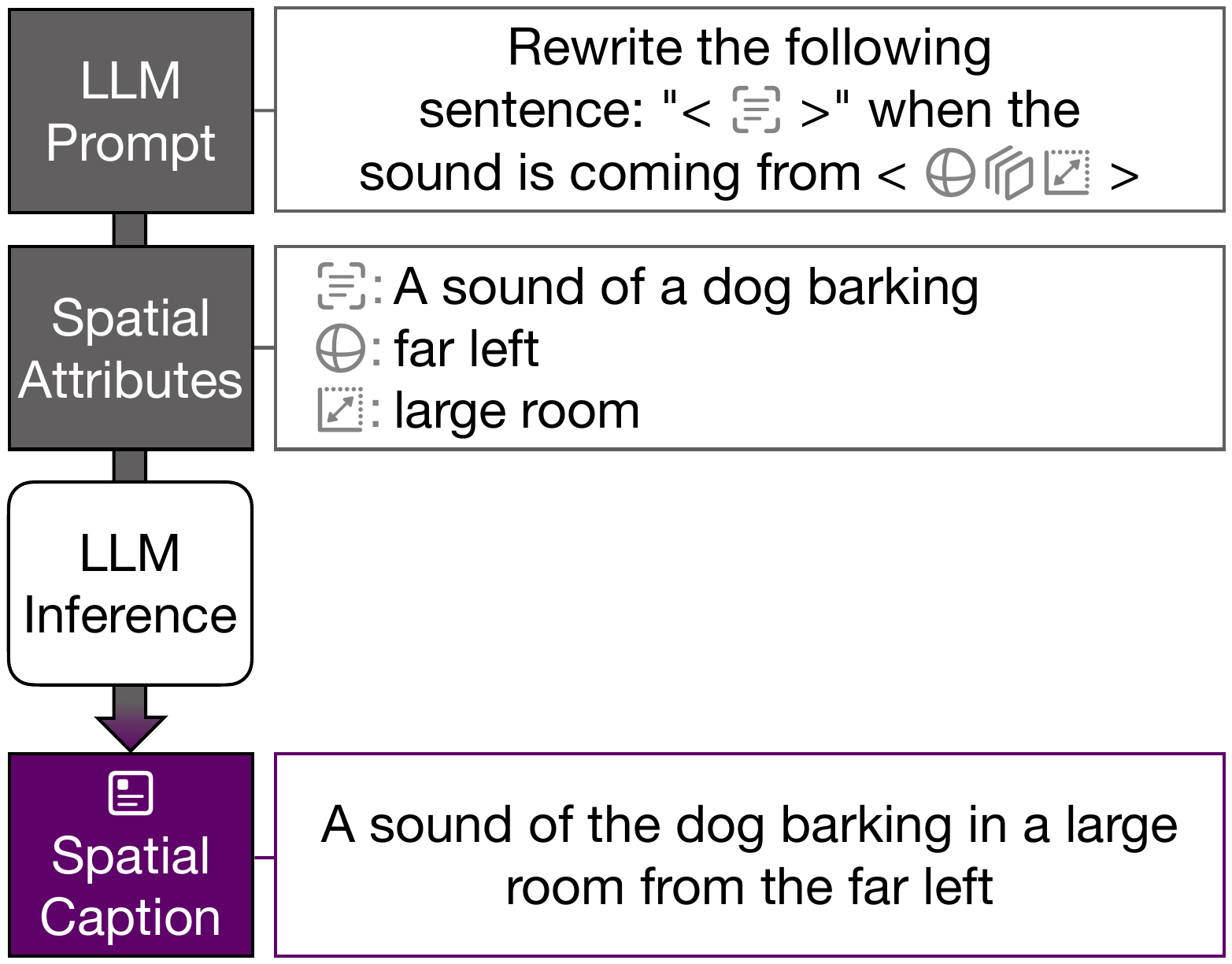}
         \caption{We augment the original captions by adding properties from the room simulations and prompt a LLM to rewrite the sentence.}
         \label{fig:llm_pipeline}
     \end{subfigure}
     \hfill
     \begin{subfigure}[t]{0.32\textwidth}
         \centering
         \includegraphics[width=\textwidth]{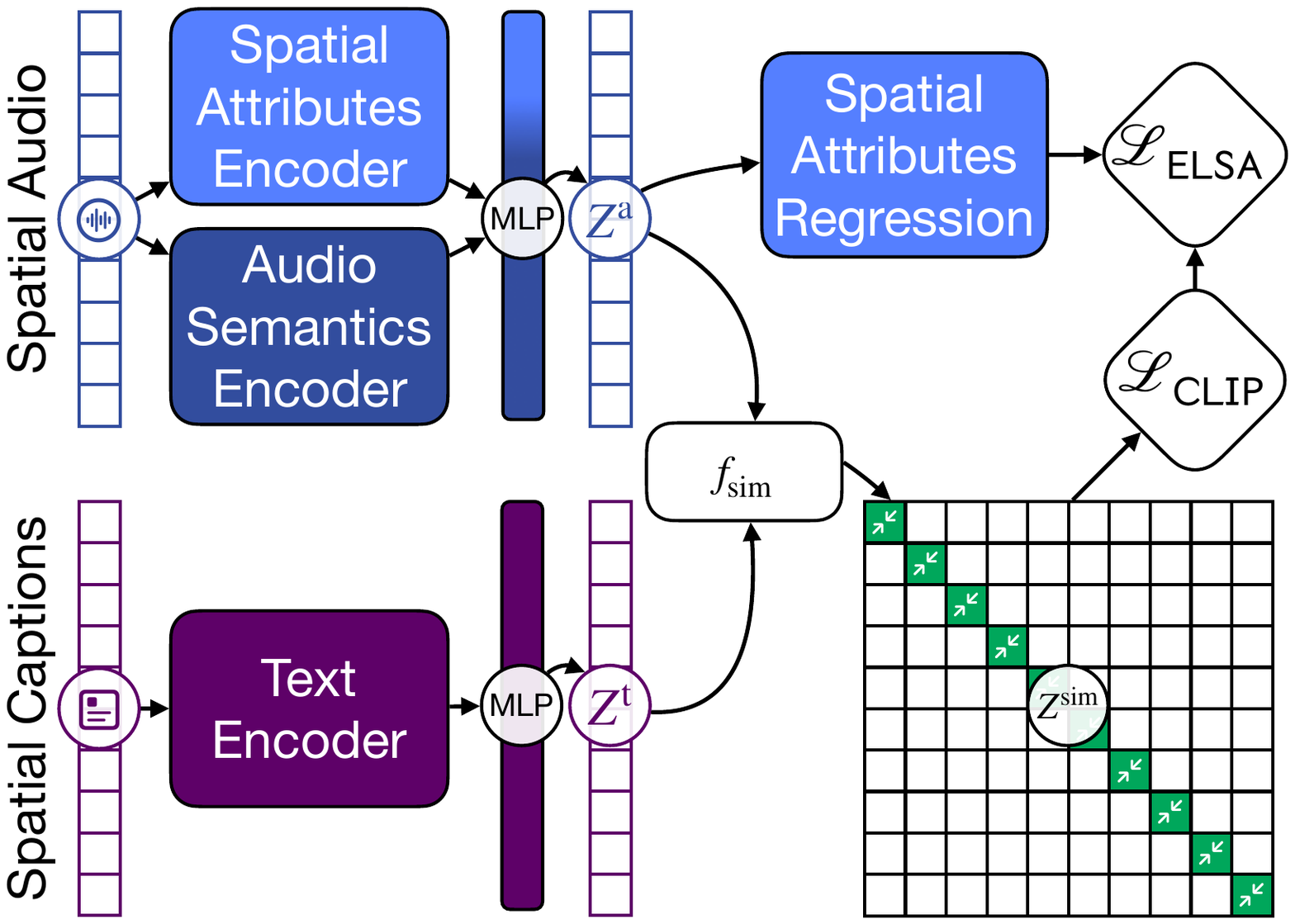}
         \caption{We encode the spatially augmented captions and audio, and then align the representations using a CLIP objective (see \cref{fig:elsa_architecture} for full architecture).}
         \label{fig:contrastive_learning}
     \end{subfigure}
        \caption{Our pipeline for learning spatial-audio representations aligned with natural language.}
        \label{fig:pipeline}
        \vspace*{-5mm}
\end{figure}

Multimodal contrastive learning approaches, e.g., CLIP~\cite{radford2021learning} and CLAP~\cite{elizalde2023clap}, use large amounts of multimodal data pairs: 413M and 634k for the LAION versions of both models~\cite{LAION400M, wu2023large}.
Training a model capable of understanding spatial audio as natural language  requires a spatial audio dataset annotated with natural language spatial descriptions (e.g., ``a dog barking in the far left corner of a room''). To the best of our knowledge, no such dataset is available. Thus, we use a spatial augmentation pipeline composed of two steps: simulating spatial audio in synthetic rooms (cf.~\cref{sec:spatial_audio_pipeline,fig:spatial_audio_pipeline}), and caption rephrasing using terms that refer to spatial audio attributes (cf.~\cref{sec:llm_pipeline,fig:llm_pipeline}). We use AudioCaps~\cite{kim2019audiocaps}, Clotho~\cite{drossos2020clotho}, and Freesound~\cite{Freesound2013} as base datasets for our augmentation pipeline.

The training set ensures at least two spatial augmentations per data point, allowing for the model to see the same audio with at least two different spatial augmentations per epoch. We generate two different sized versions of the evaluation and test sets. The larger version consists, once more, of at least two augmentations per audio sample, whilst the smaller version has no repeated samples and, consequently, is the same size as the original test set. The smaller dataset allows reporting retrieval results on the same sized dataset as the original, as size uniformity is key to consistency in retrieval metrics. The size of the respective datasets is reported in \cref{sec:dataset_statistics}.  For all datasets, we use first-order ambisonics (FOA) as the encoding of spatial audio, which we describe next.

\subsection{Spatial Audio Encoding: First Order Ambisonics}
\label{sec:ambisonics}

Monophonic, non-spatial audio, captures the spectral and temporal nature of sound, which carries a significant portion of context. Spatial audio provides additional context as it contains both spectral-temporal information and directional attributes, characterized by azimuth and elevation \(\left(\theta,\phi\right)\in \mathbb{S}^2\), and distance. Binaural audio, a common spatial audio distribution format, mimics the signal entering the ear canals. Whilst binaural audio may be a natural choice for playback over headphones, it presents challenges for encoding, storing, and processing spatial information due to the presence of head-related transfer-functions in the signal \cite{blauert2013technology}. To facilitate more processing flexibility, microphone array signals are often encoded as ambisonics~\cite{ambisonics}. This is accomplished by taking the spherical Fourier transform of the microphone signals and removing their radial component,  which is equivalent to representing the spatial signal as a phase-coincident, infinite series in a spherical harmonic basis \cite{rafaely2015fundamentals}. In practice, to avoid spatial aliasing, this series is truncated at an order proportional to the number of microphones in the array, with higher orders corresponding to a higher spatial resolution. Ambisonics are linearly mappable into a variety of audio playback formats, including binaural. First-order ambisonics (FOA) can be recorded using readily available four-channel microphone arrays, and have been shown to carry significant spatial information \cite{zotter2019ambisonics}. As such, we develop our models to ingest FOA signals. 

We leave generalization to higher orders for future work. It is worthwhile noting that once microphone array signals have been encoded into ambisonics, no a-priori knowledge on the structure of the capturing array is needed in order to perform any downstream spatial processing. Thus, ambisonics are agnostic to both recording and playback devices, making any embeddings derived from them equally generalizable.

\subsection{Spatial Augmentation of the Audio and Captions}
\label{sec:spatial_audio_pipeline}
\label{sec:llm_pipeline}

Like TUT Sounds Events 2018~\cite{adavanne2018sound} and BAT~\cite{zheng2024bat}, we use a simulator to spatially augment non-spatial audio. The augmentation pipeline mirrors that of Spatial LibriSpeech~\cite{Sarabia2023}. We specify room configurations parameterized by size, shape, and reverberation time, where reverberation time is a function of the room structure and materials with characteristic absorption and scattering coefficients. The simulator further allows specification of the placement and direction of the receiver microphones relative to the source of the sound (see \cref{fig:spatial_audio_pipeline}). For each sample we remove leading and trailing silences, and repeat the audio signal to ensure that samples are at least four seconds long before simulation.  A randomly chosen room, placement for the microphone, and placement for the static source is then selected. We ensure that the room augmentations do not overlap between the train, evaluation, and test datasets. The rooms vary in size between 13.3\sqm and 277.4\sqm, their full-band T30 reverberance ranges from 114.5ms to 2671.9ms.  The full statistics of these synthetic rooms can be found in \cref{sec:dataset_statistics}.

Our caption augmentation pipeline converts raw numerical values associated with the spatial audio attributes of the room simulator (e.g., distance to the microphone) into natural language descriptors (e.g., ``near'' or ``far''). Our caption augmentation pipeline is shown in \cref{fig:llm_pipeline}. The full mapping from spatial audio attributes to natural language is given in \cref{sec:spatial_attributes_mapping}. 

The original caption augmented with the spatial information makes up the input to LLaMA-13B~\cite{touvron2023llama}, which is prompted to rephrase in the form of a spatially augmented caption. The prompt is: 
\begin{mdframed}[style=prompt]
The sound: <\textit{original caption}> is coming from the <\textit{distance}> <\textit{elevation}> <\textit{direction}> of a <\textit{size}> <\textit{reverb}> room. Rephrase as a short English sentence describing the sound and all the details of its source.
\end{mdframed}
This template prompt overcomes challenges like non-English language in the original caption, missing spatial descriptors in the generated caption, and hallucinations that changed the meaning of the caption. We set the inference temperature of the LLM to 0.9 and the maximum tokens to 1,024. \Cref{sec:audio_dataset_captions} contains examples of the obtained spatial captions. We note that the caption re-writes can lead to hallucinations, which is discussed further in \cref{sec:hallucinations}. We leave the quantification and mitigation of hallucinations for future work.

\subsection{Spatial Real-World Dataset}
\label{sec:spatial_real_world_dataset}

Our training data consists of synthetically-augmented audio and captions, so we also recorded a small dataset to verify generalization to real-world data (refer to \cref{sec:spatial_attributes_results,sec:semantics_results} for analysis). Our spatial real-world dataset was recorded using a Zylia 19 microphone spherical array at 48kHz with a bit-depth of 24-bits per sample. The dataset contains environmental sounds typically found in an apartment. In total, we recorded 70 samples of spatial audio in five rooms. Each spatial audio sample in the dataset was captioned with the semantic content (e.g., ``sound of a vacuum''), and the direction \{``left'', ``right'', ``front'', ``back''\}, distance \{``far'', ``near''\}, and elevation \{``up'', ``down'', ``level''\}. For privacy, no personally identifiable information was included in the dataset.

\section{ELSA Pretraining for Spatial Audio and Language}
\label{sec:pretraining}

Our architecture is derived from LAION-CLAP~\cite{wu2023large}, which is composed of an audio encoder and a text encoder that aligns embeddings for similar samples across modalities whilst maintaining the original representational capabilities of the individual modalities.

\subsection{Audio Input Features}
\label{sec:ai_features}

 The audio encoder must capture both the semantics of the audio (e.g., ``the sound of a fire alarm'') and the spatial attributes (e.g., ``the upper right of a reverberant room''). Following LAION-CLAP~\cite{elizalde2023clap} and BAT~\cite{zheng2024bat}, we translate the raw audio into the frequency domain.  Consider a FOA signal represented by tensor, $\mathbf{A} \in \mathbb {C}^{ T \times F \times (N+1)^2}$, where $N=1$ is the spherical-harmonics order, $T$ the number of time frames and $F$ the number of frequency bins. More information on the derivation of $\mathbf{A}$ can be found in \cref{sec:foa_appendix}. The corresponding real-valued log-mel spectrogram feature can be written as:
\begin{equation}
    \textsc {MEL}(t,\nu) = \log \left(\left|\mathbf {A}(t,f)\right|^{2} \cdot \mathbf {W}_\text{mel}(f,\nu)\right),
\end{equation}
 where $\mathbf{W}_\text{mel}$ is the corresponding filter, $\nu$ is the filter index, $t$ is time, and $f$ is frequency.  As summarized in Table 1 of SALSA~\cite{Nguyen_2022}, both mel-spectrograms and intensity vectors (IVs) are effective spatial features for FOAs. We extract the IVs, $\boldsymbol{I}(t,f)$ as follows:
 \begin{equation}
     \label{eq:iv}
     \boldsymbol{I}_\mathrm{active}(t,f) = \Re \left[ A_\text{0,0}^{*}(t,f) \begin{pmatrix}A_\text{1,-1}(t,f)\\ A_\text{1,0}(t,f)\\ A_\text{1,1}(t,f) \end{pmatrix} \right], \quad \boldsymbol{I}_\mathrm{reactive}(t,f) = \Im \left[ A_\text{0,0}^{*}(t,f) \begin{pmatrix}A_\text{1,-1}(t,f)\\ A_\text{1,0}(t,f)\\ A_\text{1,1}(t,f) \end{pmatrix} \right],
 \end{equation}
 where $A_{n,m}$ are the $n^\textrm{th}$ and $m^\textrm{th}$ order and mode of the ambisonics signal corresponding to its omnidirectional ($W$) and three dipole $(Z,Y,X)$ components,  and $(\cdot)^*$ denotes complex conjugation. Physical normalization constants are omitted here for brevity as IVs are scaled to unit-norm \cite{Nguyen_2022}.  
 
For ELSA to use semantic features from both non-spatial audio and FOAs, during training we use sample from both the spatially-augmented datasets and the original non-spatial dataset. Since first-order ambisonics has four channels, and non-spatial audio only one, we copy the single-channel non-spatial signal across all channels. Intensity vectors normalize the dipoles by the omni channel, and result in identical IVs for non-spatial audio. We let the model learn this condition. We ablate the effect of using both spatial audio and non-spatial audio in \cref{sec:model_ablations} and find that using both improves semantic retrieval. 

\subsection{Audio and Text Encoders}
\label{sec:encoders}

Our architecture is composed of an audio encoder and a text encoder. The audio encoder consists of two branches: the \textit{semantic audio branch}, and the \textit{spatial attributes branch}. See \cref{sec:elsa_architecture} for a visualization of the full architecture.

For the semantic audio branch, we use HTSAT~\cite{chen2022htsat} since it was found to perform best in the LAION-CLAP evaluation~\cite{wu2023large}. HTSAT is a transformer-based audio encoder with self-attention blocks to achieve high performance in audio classification tasks. We initialize HTSAT with weights provided by LAION-CLAP\footnote{We use \texttt{HTSAT-fullset-imagenet-map=0.467.ckpt} from \href{https://github.com/LAION-AI/CLAP}{https://github.com/LAION-AI/CLAP}}. For spatial-audio input, we feed only the mel-spectrogram of the omni channel from the first-order ambisonics encoding. The omni channel does not contain spatial characteristics, so its role is equivalent to single channel, non-spatial audio. This branch has 30M parameters. 

As far as we are aware, there is no existing established feature encoder for spatial audio. Thus, for our spatial attributes branch we propose a two-branched CNN based on the architecture of~\cite{Sarabia2023} that was trained on a multi-task regression loss for azimuth, elevation, distance, and third-octave direct-to-reverberant ratio. The branch was trained for 100 epochs on Spatial LibriSpeech, which uses FOA spatial audio and has enough samples to train the spatial attributes branch. Further details, along with the full training hyper-parameters are discussed in \cref{sec:spatial_attributes_branch_details}. This branch is fed the active and reactive intensity vector features described in \cref{eq:iv}. This branch has 486k parameters.

The outputs of both the \textit{semantic} (768-dimensional) and the \textit{spatial attributes} (192-dimensional) branches are concatenated to form a 960-dimensional embedding. Using a two-layer multi-layer perceptron (MLP), they are subsequently projected down to a 512-dimensional embedding.

For the text branch, we follow the best performing model in LAION-CLAP~\cite{wu2023large}, and use RoBERTa-base~\cite{liu2019roberta}. RoBERTa is a general purpose bidirectional transformer~\cite{Vaswani2017}, pretrained on a dynamically masked token prediction task, which employs byte-pair encoding~\cite{Sennrich2016} for tokenization. We use the same pre-trained model as~\cite{wu2023large} as the starting point\footnote{We use the weights for \texttt{roberta-base} from: \href{https://dl.fbaipublicfiles.com/fairseq/models/roberta.base.tar.gz}{https://dl.fbaipublicfiles.com/fairseq/models/roberta.base.tar.gz}}. The text encoder has 125M parameters, and the final embedding has a dimensionality of 712, which also is projected down to 512 by a two-layer MLP, matching the size of the audio encoder output.

\subsection{Pretraining Objectives}
\label{sec:loss}

We learn aligned representations using batched contrastive loss (popularized by CLIP~\cite{radford2021learning}). The loss function rewards the alignment of representations from the same sample but different modalities, and penalizes the alignment of representations from different samples (see \cref{fig:contrastive_learning}). Our loss (in common with CLIP~\cite{radford2021learning}, CLAP~\cite{elizalde2023clap}, and LAION-CLAP~\cite{wu2023large}) is derived from the InfoNCE loss~\cite{oord2018representation}, as we now describe. Given a set of embeddings of any modality \(X \in \mathbb{R}^{N\times D}\) where the \(i^\mathrm{th}\) entry, \(x_i \in \mathbb{R}^D\) is to be matched with $y \in \mathbb{R}^D$, the following InfoNCE sample loss maximizes the similarity between the pair \(x_i\) and \(y\), and minimizes the similarity between all other \(x\) and \(y\) pairs:
\begin{equation}
\mathcal{L}_{\mathrm{InfoNCE}}(X, x_i, y) = -\log\frac{f_\mathrm{sim}(x_i, y)}{\sum_{x_j \in X} f_\mathrm{sim}(x_j, y)},
\end{equation}
where \(f_\mathrm{sim}(a, b) = \exp(a \cdot b / \tau)\) is a similarity function with a learnable temperature parameter \(\tau\). Taking the average  across all audio-text pairs in the batch, where entries at the \(i^{\mathrm{th}}\) position match each other, we arrive at the CLIP loss:
\begin{equation}
\label{eq:clip_loss}
\begin{split}
\mathcal{L}_{\mathrm{CLIP}} =& \frac{1}{2}\left(\frac{1}{N}\sum^N_{i=0}\mathcal{L}_{\mathrm{InfoNCE}}(Z^a, z^a_i, z^t_i) + \frac{1}{N}\sum^N_{i=0}\mathcal{L}_{\mathrm{InfoNCE}}(Z^t, z^t_i, z^a_i)\right), \\
= & -\dfrac{1}{2N}\sum^{N}_{i=0} \left(\log\dfrac{f_\mathrm{sim}(z^a_i, z^t_i)}{\sum^N_{j=0}f_\mathrm{sim}(z^a_j, z^t_i)} + \log\dfrac{f_\mathrm{sim}(z^t_i, z^a_i)}{\sum^N_{j=0}f_\mathrm{sim}(z^t_j, z^a_i)}\right),
\end{split}
\end{equation}

Since the rooms we use to spatially-augment the audio are parametric, we have accurate labels associated with spatial features of the audio source.  We take advantage of  these labels by adding three additional spatial regression objectives. We feed the generated 512 dimension audio embedding into three 2-layer MLPs of 33k parameters, which respectively regress the direction of arrival (azimuth and elevation) of sound in 3D space, distance of the source to the receiver, and room floor area. These objectives, along with the CLIP loss in \cref{eq:clip_loss}, define our final loss:
\begin{equation}
    \mathcal{L}_\mathrm{ELSA} = \mathcal{L}_\mathrm{CLIP} + \mathcal{L_\mathrm{dir}} +  \mathcal{L_\mathrm{dist}} +  \mathcal{L_\mathrm{area}},
\end{equation}
where \(\mathcal{L}_\mathrm{dir}\) is the cosine similarity between the predicted and target angles, and  \(\mathcal{L}_\mathrm{dist}\) and \(\mathcal{L}_\mathrm{area}\) is the mean-squared error between the predicted and target distances and room floor area respectively. We ablate the differences between \(\mathcal{L}_\mathrm{ELSA}\) and \(\mathcal{L}_\mathrm{CLIP}\) in  \cref{sec:model_ablations} and find that at a negligible cost (0.4\%) to semantic retrieval, we get a 15.3\% improvement to 3D localization capability and 12.3\% improvement in distance estimation when using \(\mathcal{L}_\mathrm{ELSA}\).

\section{Experiments, Results, and Discussion}
\label{sec:experiments}

\begin{table}[t]
  \scriptsize
  \caption{Comparison of model capabilities and performance for retrieval of semantic captions from AudioCaps, and 3D sound localization for the REAL component TUT Sound Events 2018. ELSA is the only model that allows both open vocabulary language understanding and spatial localization, and performs comparably against the baselines for both tasks.}
  \label{tab:spatial_audio_regression}
  \vspace*{2mm} 
  \centering
  \setlength{\tabcolsep}{1mm}
  \begin{tabular}{
    >{\raggedright\arraybackslash}m{30mm}%
    >{\centering\arraybackslash}m{18mm}%
    >{\centering\arraybackslash}m{18mm}%
    >{\centering\arraybackslash}m{18mm}%
    >{\centering\arraybackslash}m{18mm}%
  }
    \toprule
    \textsc{Model} & \textsc{Semantic Capabilities} & \textsc{Spatial Capabilities} & \textsc{AudioCaps mAP@10}$\uparrow$ & \textsc{REAL 3D Local.}(\mdeg)$\downarrow$\\
    \midrule
    SeldNET \cite{adavanne2018sound} & \xmark~Limited vocab. & \cmark & \xmark & 26.6 
    \\ 
    PILOT \cite{schymura2021pilot} & \xmark~Limited vocab. & \cmark & \xmark & 4.2 \\
    \midrule
    Spatial Librispeech \cite{Sarabia2023} & \xmark & \cmark & \xmark & 12.4\\
    LAION-CLAP \cite{wu2023large}  & \cmark~Open vocab. & \xmark &  43.8 & 95.29\\
    ELSA (ours) & \cmark~Open vocab. & \cmark & \textbf{44.2} & 14.97\\
    \bottomrule
  \end{tabular}
  \vspace*{-3mm}
\end{table}

We demonstrate that ELSA jointly captures the semantics and spatial attributes of sound with either audio or text inputs by answering the following research questions:

\begin{description}
    \item[RQ1] Does ELSA capture \emph{spatial attributes} in spatial audio (\cref{sec:spatial_attributes_results})?
    \item[RQ2] Does ELSA capture \emph{semantic information} in both text and audio (\cref{sec:semantics_results})?
    \item[RQ3] Does ELSA transfer to our real-world dataset? (Sections \ref{sec:spatial_attributes_results} and \ref{sec:semantics_results})?
    \item[RQ4] Does ELSA provide interpretable multimodal representations (\cref{sec:embedding_structure})?
    \item[RQ5] Are ELSA embeddings capable of driving automatic captioning (\cref{sec:caption_gen})?
\end{description}

\begin{wraptable}{r}{70mm}
  \vspace*{-6mm}
  \scriptsize
 \caption{
 Zero-shot classification accuracy using the cosine similarity between test set audio embeddings and templated probe caption embeddings. The template is ``\texttt{A sound coming from <spatial attribute>}'' and a value for <spatial attribute> is substituted into the template representing the desired class (e.g., ``near'' or ``far'' for distance). A classification is correct if the attribute in the closest test sample matches the attribute in the template. We cannot provide comparisons with baselines since this is a new task.
 }
  \label{tab:spatial_attributes_retrieval}
  \vspace*{2mm}
  \centering
  \begin{tabular}{lccc}
    \toprule
    \textsc{Task} & \textsc{S-Clotho} & \textsc{S-AC} & \textsc{S-RWD} \\
    \midrule
    Distance (2-class)       & 96.0\%  & 92.9\% & 67.1\% \\
    Direction (4-class)      & 92.0\%  & 92.8\% & 35.8\% \\
    Elevation (2-class)      & 100.0\%  & 100.0\% & 72.1\% \\
    Room area (2-class)      & 76.6\%  & 74.7\% & N/A \\
    Reverberation (2-class)  & 100.0\%   & 83.3\% & N/A \\
    \bottomrule
  \end{tabular}
  \vspace*{-10mm}
\end{wraptable}

\subsection{Training and Evaluation of ELSA}

As indicated in \cref{sec:encoders}, we use pretrained weights for the semantic audio encoder, the spatial attributes encoder, and the text encoder. All components of the model are fine-tuned, which corresponds to 158M trainable parameters, an increase of 0.86\% over LAION-CLAP\cite{chen2022htsat}.

For our best model, we train for 40 epochs on 12 nodes, each with 8 NVIDIA A100 GPUs and $96$ CPU cores with a batch size of 2,304. Training converges within 17 hours. We use the Adam optimizer with a learning rate of \(5\times10^{-5}\) and cosine scheduling. We select the checkpoint with the lowest mAP@10 retrieval on the spatially augmented captions.

\subsection{Spatial Attributes Evaluation}
\label{sec:spatial_attributes_results}
We show that ELSA captures the spatial attributes of sound (\textbf{RQ1}) by carrying out downstream regression and zero-shot spatial prompt classification.  For regression to 3D sound localization, we a train two-layer MLP with 32,768 parameters using the ELSA audio embeddings generated from the training set. We then evaluate on the REAL component of the TUT Sound Events 2018 dataset~\cite{adavanne2018sound}.  \Cref{tab:spatial_audio_regression} confirms that CLAP cannot encode spatial attributes (95.29\mdeg), whereas ELSA achieves 14.97\mdeg mean-absolute error (MAE) and maintains a higher mAP@10 for semantic retrieval tasks than CLAP. \cref{sec:doa_error_analysis} shows that there is little variability in the direction-of-arrival error across various spatial attributes. However, we note the errors tend to be higher at the extrema of the dimensions. When compared to methods designed explicitly for 3D sound localization, ELSA performs better than SeldNET  (the baseline included with TUT Sound Events 2018) by +11.6\mdeg, and that achieves only -2.6\mdeg MAE compared to the model in Spatial LibriSpeech\footnote{The Spatial LibriSpeech model can be considered a supervised version of ELSAs contrastive learning, since the authors also train on a synthetically augmented dataset.}. ELSA does not reach the performance of PILOT (4.3\mdeg), but this model was specifically-tuned only for 3D sound localization on data derived from TUT Sound Events 2018 \cite{schymura2021pilot}.

To verify that the spatial attributes are aligned with language, we create new captions using the template: ``\texttt{A sound coming from <\textit{spatial attribute}>}'', where \texttt{<\textit{spatial attribute}>} can be distance, direction, elevation, room size, and reverberation. For instance, a caption for distance might be ``\texttt{A sound coming from far away}''. The ELSA text embeddings for such captions are extracted from the pre-trained encoder and compared in a zero-shot fashion with ELSA audio embeddings for samples from the test set using cosine similarity.  We classify the match as correct if the spatial attribute in the closest audio sample matches the spatial attribute of the query caption, and we report accuracy in  \cref{tab:spatial_attributes_retrieval}.  ELSA achieves >90\% correct retrieval for most spatial attributes. For room area, ELSA achieves 74.7\% correct retrieval, which we hypothesize is due to the relatively small perceptual differences between small (<50\sqm) and large rooms (>100\sqm). We observe a transfer gap on retrieval scores when evaluating on our spatial real-world dataset, with ELSA achieving 67.1\% (distance), 35.8\% (direction) and 72.1\% (elevation) correct retrieval. Part of this performance difference is because the spatial attributes were only estimates by the annotators during data capture. On the other hand, performance using pre-trained LAION-CLAP is close to random for all tasks (\cref{sec:spatial_attributes_clap}), which is expected as CLAP was not trained with spatial audio or captions.

\subsection{Semantics Evaluation}
\label{sec:semantics_results}
\begin{table}[t]
  \scriptsize
  \caption{Semantic retrieval (R@1, R@5, and R@10) for CLAP and ELSA calculated over the original (non-spatial) versions of Clotho and AudioCaps.  Although ELSA is trained using a mixture of non-spatial and spatial audio, it conserves the retrieval performance on non-spatial audio of LAION-AI CLAP, which was trained on only non-spatial data.  For the training data, read \(C\) as Clotho, \(AC\) as AudioCaps, \(LA\) as LAION-Audio-630K and \(FS\) as Freesound.  A superscript \(^{S}\) denotes the spatially-augmented equivalent dataset. We use Freesound, a subset of LAION-Audio-630K due to its more permissive licensing. For a fair comparison, we train a version of CLAP locally with Clotho, Audiocaps and Freesound, which is not reported in the CLAP paper.}
  \label{tab:semantic_retrieval_mono}
  \vspace*{2mm} 
  \centering
  \setlength{\tabcolsep}{2pt} 
  \begin{tabular}{ll @{\hskip 3mm} llllll @{\hskip 3mm} llllll}
    \toprule
    \multicolumn{2}{c}{} & \multicolumn{6}{c}{\textsc{AudioCaps}} & \multicolumn{6}{c}{\textsc{Clotho}}                   \\
    \multicolumn{2}{c}{} & \multicolumn{3}{c}{\textsc{Text-to-Audio}} & \multicolumn{3}{c}{\textsc{Audio-to-Text}} & \multicolumn{3}{c}{\textsc{Text-to-Audio}} & \multicolumn{3}{c}{\textsc{Audio-to-Text}} \\
    \cmidrule(r){3-14}
    \textsc{Model} & \textsc{Train Data}  & R@1 & R@5 & R@10 & R@1 & R@5 & R@10 & R@1 & R@5 & R@10 & R@1 & R@5 & R@10 \\
    \midrule
    CLAP (paper) & \(C,AC,LA\) & 34.7  & 70.5 & 83.2 & 45.3 & 79.5  & 89.2 & 16.4 & 39.0 & 51.0 & 21.8 & 44.6 & 60.1\\
    \midrule
    CLAP (local) & \(C,AC,FS\) & 32.7  & 68.8  & 81.5  & 40.7 & 74.0  & 84.7 & 14.4 & 37.6 & 50.7 & 18.3 & 40.5 & 55.1 \\ 
    ELSA & \(C,AC,FS,C^{S},AC^{S},FS^{S}\) & \textbf{33.2} & 68.2 & 81.0 & \textbf{40.9}  & \textbf{74.4}  & \textbf{86.1}  & \textbf{15.0} & 36.7 & \textbf{50.8} & \textbf{20.1} & \textbf{43.2} & \textbf{55.4}\\   
    \bottomrule
  \end{tabular}
  \vspace*{-3mm}
\end{table}
\raggedbottom

Following LAION-CLAP~\cite{wu2023large}, we calculate retrieval results when finding matches from audio-to-text and text-to-audio. To compute retrieval, we encode the test set for each modality, and for every sample we check whether the corresponding sample in the other modality has the closest cosine distance (R@1), is within the five closest samples (R@5), or within the ten closest samples (R@10).  The results in \cref{tab:semantic_retrieval_mono} show that in addition to learning representations of spatial captions and spatial audio, ELSA also performs on par with LAION-CLAP on non-spatial tasks.  \Cref{tab:semantic_retrieval_spatial} in \cref{sec:semantic_retrieval_appendix} shows the retrieval results when using spatially-augmented versions of AudioCaps and Clotho. We remark that adding Freesound to the training set decreases the retrieval scores in Spatial AudioCaps, but improves retrieval scores in Clotho, due to Clotho being a differently-captioned subset of Freesound. We note that the spatial retrieval performance of ELSA is lower than the non-spatial retrieval performance (for instance, -9.4\% and -13.3\% on audio-to-text and text-to-audio R@10 AudioCaps). This reflects the fact that spatial captions are harder to match, since there is a larger number of attributes and since there are hard-negatives (same semantics, different spatial attributes). Still, ELSA achieves the highest retrieval scores on the spatial real-world dataset (\cref{tab:semantic_retrieval_spatial_rwd} in \cref{sec:semantic_retrieval_appendix}), showcasing its ability to transfer to the real-world without fine-tuning. Note that we cannot provide comparisons with prior models since using these spatial augmentations is a new task.

\subsection{Interpreting the representation structure of ELSA}
\label{sec:embedding_structure}

To confirm that directional characteristics in ELSA spatial caption embeddings are encoded in the same feature space as those in ELSA spatial audio embeddings, we train a  direction regressor with a two-layer MLP using the \textit{spatial audio embeddings} in the training split. We subsequently regress the \textit{spatial text embeddings} to azimuth values using our trained regressor and affix direction labels (``left'', ``right'', ``front'', ``back'') to each sample based on the azimuth values. We obtain 64.3\% accuracy on the four-class problem, indicating alignment between the encoding of the modalities. Similarly, we obtain an accuracy of 76.5\% for when classifying over distance labels (``far'', ``near'') and 55.1\% over elevation labels (``up'', ``down'').
\begin{wrapfigure}{r}{65mm}
    \centering
    \includegraphics[width=64mm, trim={3mm 10mm 3mm 10mm},clip]{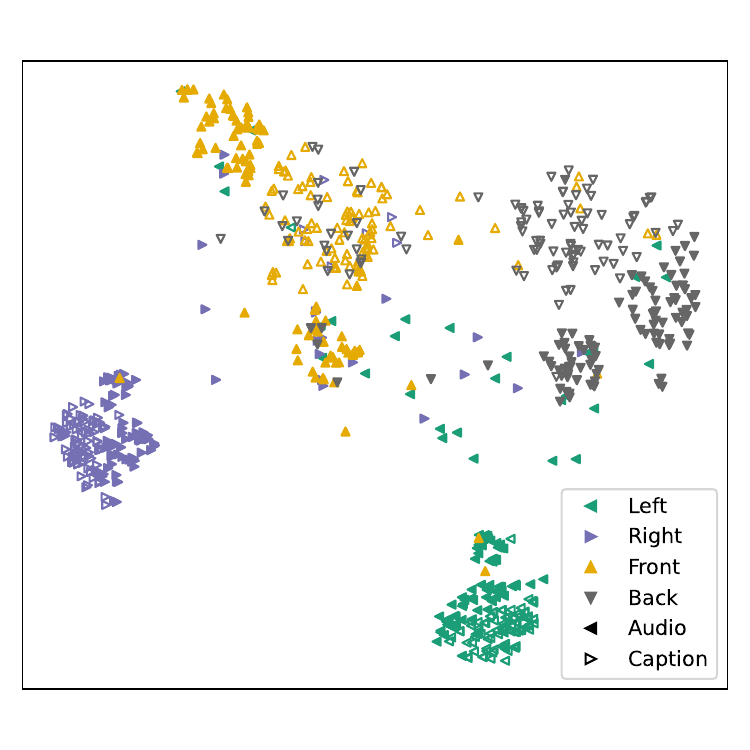}

    \caption{UMAP projection of ELSA embeddings of the test splits of Spatial-Clotho and Spatial-AudioCaps. Filled markers are obtained from spatial audio, and hollow markers are obtained from spatial captions. The UMAP projection was fitted with the train splits of Spatial-Clotho and Spatial-Audio caps, and we made use of supervised dimension reduction to highlight the \textit{direction} differences rather than the semantic differences in the embeddings.}
    \label{fig:direction_cluster}
    \vspace*{-5mm}
\end{wrapfigure}

Besides using regression to confirm  that ELSA embeddings capture spatial direction, we verify whether the ELSA embeddings can be clustered by spatial attributes. \cref{fig:direction_cluster} shows a UMAP projection of the ELSA embeddings from the test sets of Spatial-AudioCaps and Spatial-Clotho. Note that the UMAP projection was guided with the embeddings and labels of the training sets of both datasets. The figure shows the embeddings cluster well with the direction labels, though there is some degree of confusion between ``back'' and ``front''. This is corroborated by the analysis in \cref{sec:embedding_clusters}, where we compute Wassertein distances directly in the 512-dimensional space. We carried out a similar analysis for spatial distance and found the embeddings cluster clearly between ``near'' and ``far''.

We validate that ELSA audio embeddings capture implicit spatial attributes that are latent in the text-encoder by first training a classifier using the spatial audio in our training data, where the classes are broad directions, such as \emph{above} and \emph{below}.  We use LLaMA-13B~\cite{touvron2023llama} to generate descriptions of sounds that would typically come from each of these directions, e.g.,  ``the rhythmic drumming of raindrops on a skylight'' (for above) and ``the faint creaking of an old house settling'' (for below). \cref{implicit_texts} lists all generated captions. Finally, we use the classifier trained on audio samples to classify ELSA embeddings for these generated captions with implicit directionality. We find that the classifier can correctly identify 68\% of the sounds typically heard from above as being from above, showing that the latent space of the text encoder for ELSA is capturing directionality.

Lastly, we show that we can swap the spatial direction encoded by an ELSA audio embedding with a simple text caption. We first obtain ELSA prototypes for four directions (``left'', ``right'', ``front'', ``back'') with the template: ``\texttt{A sound coming from the \textit{direction}}''. Next, we train a 4-class direction classifier with a two-layer MLP using the spatial audio in the training splits of our spatially-augmented datasets. To swap the direction of the sound, we subtract the text prototype of the original  direction and add prototype for the new direction. For evaluation, we swap the spatial direction of every sample in our spatially-augmented test set that was correctly classified by the 4-class direction classifier (96.7\% of the audio embeddings). Our results show that 99.7\% of the swapped samples are classified correctly with the new spatial direction, which highlights the strong alignment of spatial features across modalities, resulting in the ability to edit spatial attributes of existing spatial audio using text in embedding space. Further details about this experiment are described in \cref{sec:direction_transposition}. These results also point to exciting avenues wherein text can condition the manipulation and generation of spatial characteristics of audio. We leave this application for future work.

\subsection{Spatial Audio Caption Generation}
\label{sec:caption_gen}

\begin{wraptable}{r}{60mm}
  \vspace*{-6mm}
  \scriptsize
  \caption{
    Evaluation of Spatial Audio Caption Generation. Metrics were obtained from the Audio Captioning task of the DCASE Challenge\protect\footnotemark~by comparing the generated captions produced from spatial audio and the ground-truth captions from the test splits of Spatial-AudioCaps (S-AC) and Spatial-Clotho.
  }
  \label{tab:caption_generation_metrics}
  \vspace*{2mm}
  \centering
  \begin{tabular}{lccc}
    \toprule
    \textsc{Metric} & \textsc{Range} & \textsc{S-Clotho} & \textsc{S-AC} \\
    \midrule
    SPIDEr~\cite{SPIDEr}          & [0, 5.5]       & 0.19 & 0.34 \\
    FENSE~\cite{FENSE}            & [-1.0, +1.0]   & 0.59 & 0.68 \\
    \# Unique words               & [0, \(\infty\))  & 1103 & 1258 \\
    \bottomrule
  \end{tabular}
\end{wraptable}

\footnotetext{Available at \href{https://github.com/Labbeti/aac-metrics}{https://github.com/Labbeti/aac-metrics}.}

Decoding multimodal embeddings into natural language can be achieved by prefixing an autoregressive causal language model~\cite{clipcap, gu2023, kim2023, deshmukh2024} , where the prefix is constructed from a projection of the multimodal embeddings. To facilitate audio captioning using ELSA, we fine-tune a GPT-2 model~\cite{gpt2} with 12 attention layers each having 12 heads (with 163M parameters). The ELSA embeddings are projected onto the prefix using a single dense layer (393k parameters). With the ELSA encoder frozen, we train the GPT-2 model on 150k spatial-audio embedding and caption pairs from Spatial-Clotho and Spatial-AudioCaps. We report caption generation metrics in \cref{tab:caption_generation_metrics} and show three generation samples in \cref{sec:caption_gen_examples}. Overall, we find that automatic spatial audio captioning systems are viable though more work is needed to increase the vocabulary size of the generations.

\section{Conclusions, Limitations, and Further Work}

We have presented ELSA, an AFM that aligns representations of spatial audio and equivalent text descriptions. To train such representations we built a pipeline to spatially augmented the audio in existing non-spatial audio-text datasets, such as Clotho~\cite{drossos2020clotho} and AudioCaps~\cite{kim2019audiocaps}, and added spatial information to their respective captions. Our results show that ELSA embeddings capture \emph{both} the semantic contents and the spatial attributes of the audio, with ELSA achieving +2.8\% higher scores in audio-to-text and text-to-audio retrieval scores than the  state-of-the-art, and obtaining -2.6\mdeg MAE in direction of arrival error with respect to an equivalent baseline. Interestingly, by mixing spatial and non-spatial audio and caption pairs, ELSA is able to represent non-spatial audio as well, Finally, we show that the representation space of ELSA is structured in that the directionality of a spatial audio sample can be transposed by simple addition or subtraction of two text representations. Future work will explore acoustic scenarios with overlapping sound sources and sound sources that are moving in the scene. ELSA will also benefit from advances in spatial attributes encoders. In this work, we used the augmented spatial captions as is, but further work should ensure consistency with the semantics before and after augmentation, which will further improve the representational power of ELSA.

Perceiving spatial audio is a fundamental aspect of human nature. As is linking perception with language. Using spatial audio and a contrastive multimodal training approach, ELSA bridges the gap between feature rich spatial audio and language, paving the way for more intuitive and effective human-machine interactions by allowing for richer understanding of the users' environment and generation of immersive sound scenes from natural language. 

\paragraph{Broader Impact} Our research has the potential to be used in creation of immersive augmented or virtual reality environments. If not controlled well, these immersive experiences have the potential to become addictive, and thus impact the mental health of individuals or even society as a whole. Another danger is possibility of creating deepfakes of soundscapes, thus making it possible for generated 3D environments to sound very realistic. The proliferation of deepfake soundscapes could lead to misinformation and manipulation, undermining trust in audio media.

\newpage

\section*{Acknowledgments}
The authors would like to thank Nicholas Apostoloff, Masha Fedzechkina, Rin Metcalf, Russ Webb, Megan Maher Welsh, and Luca Zappella for their insightful input and discussions on earlier versions of this paper. Moreover, we are thankful to Denise Hui and David Koski for technical support. Names are in alphabetical order by last name within group.

\printbibliography

\newpage

\appendix
\counterwithin*{figure}{part}
\stepcounter{part}
\renewcommand{\thefigure}{A.F.\arabic{figure}}

\counterwithin*{table}{part}
\stepcounter{part}
\renewcommand{\thetable}{A.T.\arabic{table}}

\section{Appendix}

\subsection{Dataset statistics}
\label{sec:dataset_statistics}

\cref{tab:spatial_dataset_sources} presents a summary of all the paired audio and text datasets we use for training and evaluation.

\begin{table}[h]
  \scriptsize
  \caption{Audio-caption dataset descriptions.  The first three rows correspond to the original publicly available datasets, and the subsequent rows correspond to our spatially-augmented variants. For each spatially augmented dataset, there are at least two spatial augmentations per original sample in the train split.}
  \label{tab:spatial_dataset_sources}
  \vspace*{2mm}
  \setlength{\tabcolsep}{1mm}
  \centering
  \begin{tabular}{
    >{\raggedright\arraybackslash}m{20mm}%
    >{\centering\arraybackslash}m{10mm}%
    >{\centering\arraybackslash}m{15mm}%
    >{\centering\arraybackslash}m{15mm}%
    >{\centering\arraybackslash}m{15mm}%
    >{\centering\arraybackslash}m{50mm}%
  }
    \toprule
    \textsc{Dataset} & \textsc{Spatial Audio} & \textsc{Splits} & \textsc{Num. Samples} & \textsc{Duration (hrs)} & \textsc{Caption Description}\\
    \midrule
    Clotho & \ding{55} & train, val, test & 3,839 & 23.99 & 5 captions per audio\\
    AudioCaps &  \ding{55} & train, val, test & 49,274 & 136.87 & 1--2 captions per audio\\
    FreeSound & \ding{55}  & train, val, test & 414127 & 2,528.15 & 1--2 captions per audio, keyword tags\\
    Spatial-Clotho  & Synthetic  & train, val, test & 8,546 & 55.0 &  5 spatially augmented captions per audio\\
    Spatial-AudioCaps  & Synthetic   & train, val, test & 98,459 & 258.12 & 1--2 spatially augmented captions per audio\\ 
    Spatial-FreeSound  & Synthetic  & train, val, test & 783,033 & 4,425.53 & 1--2 spatially augmented captions per audio \\
    Spatial-RWD & Recorded & test & 70 & 0.25 & 1--2 human annotated spatial captions per audio\\
    \bottomrule
  \end{tabular}
\end{table}

For the spatially-augmented versions of Clotho, AudioCaps, and Freesound, we use 8,972 parametric rooms with the statistic described in \cref{tab:rsim_statistics}. Note that the parametric rooms in the test set are a subset of the rooms in the training set, however the sources locations on those rooms do not overlap.

\begin{table}[h]
    \scriptsize
    \centering
    \setlength{\extrarowheight}{1mm}
    \setlength{\tabcolsep}{2pt} 
    \caption{Spatial attributes of room simulations used to spatially-augmented audio and language datasets}
    \label{tab:rsim_statistics}
    \vspace*{2mm} 
    \begin{tabular}{
    >{\raggedleft\arraybackslash}m{0.20\columnwidth}%
    >{\centering\arraybackslash}m{0.20\columnwidth}%
    >{\centering\arraybackslash}m{0.20\columnwidth}%
    }
    \toprule
    
    & 
    \textsc{Train} \&
    \textsc{Validation} &
    \textsc{Test} \\
    
    \midrule
    
    Number of simulated rooms & 8,952 & 4,970 \\
    
    Source azimuth & %
        [-180.0\mdeg, +180.0\mdeg] & %
        [-180.0\mdeg, +180.0\mdeg]\\
    
    Source elevation & %
        [-47.5\mdeg, +48.7\mdeg] &
        [-29.8\mdeg, +42.4\mdeg] \\
    
    Source distance & %
        [0.5m, 4.0m] &%
        [0.9m, 4.0m] \\
    
    Room floor area & %
        [13.3\sqm, 277.4\sqm] &
        [14.3\sqm, 277.4\sqm] \\

    Full-band T30 & %
        [144.5ms, 2671.9ms] &
        [167.8ms, 1254.8ms]\\
     
    \bottomrule
    \end{tabular}
\end{table}

\subsection{Mapping of spatial attributes to natural language}
\label{sec:spatial_attributes_mapping}
As part of the spatial-augmentation pipeline (described in \cref{sec:llm_pipeline}), we use the mappings in \cref{tab:spatial_audio_mapping} to convert spatial attributes to natural language.

\begin{table}[h]
    \scriptsize
    \centering
    \caption{Mapping between spatial features and natural language descriptors}
    \label{tab:spatial_audio_mapping}
    \vspace*{2mm} 
    \setlength{\tabcolsep}{2pt} 
    \begin{tabular}{
    >{\raggedleft\arraybackslash}m{0.20\linewidth}%
    >{\centering\arraybackslash}m{0.20\linewidth}%
    >{\centering\arraybackslash}m{0.20\linewidth}%
    >{\raggedright\arraybackslash}m{0.22\linewidth}%
    }
      \toprule
      \textsc{Spatial Feature} & \textsc{Range} & \textsc{Bounds} & \textsc{Language Descriptor} \\
      \midrule
      \multirow{2}{*}{Distance}  & \multirow{2}{*}{[0m, 5m]}      & < 1 m    & \textit{near} \\
                               &                                    & > 2 m    & \textit{far} \\
      \midrule
      \multirow{4}{*}{Direction}     & \multirow{4}{*}{[-180\mdeg, +180\mdeg]}  & [-55\mdeg, -125\mdeg] & \textit{left} \\
                                     &                              & [+55\mdeg, +125\mdeg] & \textit{right} \\
                                     &                              & [-35\mdeg, +35\mdeg] & \textit{front} \\
                                     &                              & [-145\mdeg, +45\mdeg] & \textit{back} \\
      \midrule
      \multirow{2}{*}{Elevation}     & \multirow{2}{*}{[-48.1\mdeg, +48.7\mdeg]}    & > 40.0\mdeg & \textit{up} \\
                                     &                              & < -40.0\mdeg & \textit{down} \\
      \midrule
      \multirow{2}{*}{Reverberation} & \multirow{2}{*}{[144.5ms, 2671.9ms]}    & > 1000ms & \textit{highly reverberant} \\
                                     &                              & < 200ms & \textit{acoustically dampened} \\
      \midrule
      \multirow{2}{*}{Room Floor size}     & \multirow{2}{*}{[13.3\sqm, 277.4\sqm]}    & < 50\sqm & \textit{small} \\
                                     &                              & > 100\sqm & \textit{large} \\
      \bottomrule
    \end{tabular}
\end{table}

\subsection{Audio dataset captions}
\label{sec:audio_dataset_captions}

We report the mapping from raw spatial values to spatial captions in \cref{sec:spatial_attributes_mapping}. We select these bounds based on how an audio would be percieved by human ears, e.g. a sound higher than 40 degrees elevation sounds like its coming from a height. 

We also present a few random samples of regular to spatial text rewrites by the LLM:

\begin{captionenv}[frametitle={Inputs to LLM}]
\begin{description}
    \item[Original caption] A bird is loudly making a lot of noises.
    \item[Distance] far
    \item[Room Size] medium
\end{description}
\mdfsubtitle{Rewritten caption}
In a medium-sized room, a bird is emitting loud sounds from a
distant location.
\end{captionenv}

\begin{captionenv}[frametitle={Inputs to LLM}]
\begin{description}
    \item[Original caption] A bunch of people are talking in the background while a man talks to another.
    \item[Room size] medium
\end{description}
\mdfsubtitle{Rewritten caption}
In a medium-sized room, the sound of multiple
people conversing in the background can be heard, alongside a man speaking to
another individual.
\end{captionenv}

\begin{captionenv}[frametitle={Inputs to LLM}]
\begin{description}
    \item[Original caption] A motor vehicle is running with speed and stopped its engine.
    \item[Distance] far
    \item[Reverb] highly reverberant
\end{description}
\mdfsubtitle{Rewritten caption}
The sound of a motor vehicle running at high speed and then abruptly stopping its engine is emanating from the far end of a highly reverberant room.
\end{captionenv}

\begin{captionenv}[frametitle={Inputs to LLM}]
\begin{description}
    \item[Original caption] A fire crackles in the background while a faint knocking fades away over time.
    \item[Direction] front
    \item[Distance] far
    \item[Room size] small
\end{description}
\mdfsubtitle{Rewritten caption}
The sound of a fire crackling and a faint knocking can be heard emanating from the far end of a small room.
\end{captionenv}

\begin{captionenv}[frametitle={Inputs to LLM}]
\begin{description}
    \item[Original caption] A distant voice and chirping while the wind blows strongly.
    \item[Direction] right
    \item[Distance] near
    \item[Room size] large
\end{description}
\mdfsubtitle{Rewritten caption}
In a large room, a sound of a distant voice and chirping is emanating from the right side, and it is accompanied by the strong wind blowing nearby.
\end{captionenv}

\subsection{Hallucinations in LLMs}
\label{sec:hallucinations}

We note that the caption re-writes can lead to hallucinations in captions. For instance, the phrase \texttt{``the purr of a carerra''} describing the sound of the engine of a Porsche Carerra was rephrased as \texttt{``the purring cat named Carerra''}. Another example, \texttt{``the sounds of papers turning''} with direction \textit{below} was rewritten as \texttt{``the sound of someone shuffling cards in the basement''}. It is not clear what is the overall effect of the hallucinations; as in the first example it changes the semantics of the audio, but in the second example adds plausible and welcome diversity to our caption set. As mentioned in the main section, we leave the quantification and mitigation of sub optimal hallucinations for future work.

\subsection{Further background on first-order ambisonics}
\label{sec:foa_appendix}

Consider a continuum of plane-waves impinging on the surface of a sphere, $p(kr,\Omega)$, where $p$ is acoustic pressure, $k=2\pi f c^{-1}$ is the spatial frequency, $r$ is radial distance and $\Omega\equiv(\theta,\phi)\in \mathbb{S}^2$ is direction in terms of elevation $\theta$ and azimuth $\phi$. The expansion of this function in a spherical harmonics basis, $p_{nm}(k)$ can be written as,
\begin{equation}\label{eqn:fwd_sft}
\begin{split}
    p_{nm}(k,r) &= {b_n}(kr)\int_{\Omega \in {S^2}} a(k,\Omega){\left[{Y_n^m(\Omega)} \right]^\ast}{\rm d}\Omega \cr &= {b_n}(kr){A_{nm}}(k),
\end{split}
\end{equation}
Here, $a(k,\Omega)$ denotes the plane-wave density function in the spatial domain, $Y_n^m(
\Omega)$ are the spherical-harmonics basis function for order $n$ and mode $m$, and $b_n(kr)$ is the radial function given for a rigid sphere by \cite{williams1999fourier}, 
\begin{equation}
    b_n(kr)=4 \pi i^{n} \left [j_{n}(kr)- \frac{{j^{\prime}_{n}(kr_{0})}} {h^{\prime}_{n}(kr_{0})} h_{n}(kr) \right]
\end{equation}
where $j_n(kr)$ and $h_n(kr)$ are the spherical Bessel and Hankel function, respectively, and $(\cdot)'$ denotes their first derivative with respect to the argument. In this work we employ a real-valued spherical harmonics basis and radial functions corresponding to a rigid sphere. We denote $A_{nm}(k)$ as the spherical Fourier transform of the plane-wave density function, $a(k,\Omega)$, and refer to it as an ambisonics signal. We further denote the inverse spherical Fourier transform of the ambisonics signal as,
\begin{equation}\label{eqn:bkwd_sft}
    a(k,\Omega)=\sum_{n=0}^{\infty}\sum_{m=-n}^{n}A_{nm}(k)Y_{n}^{m}(\Omega),
\end{equation}

Considering now a microphone array with \(Q\) sensors, the integral in \eqref{eqn:fwd_sft} becomes a weighted finite summation. In order to avoid spatial aliasing, we conform to \(Q=(N+1)^2\) with optimal spatial sampling of \(Q\) \cite{rafaely2004analysis} sensors, where \(N\) is the spherical harmonics order. Accordingly the outer sum in \eqref{eqn:bkwd_sft} is truncated at order \(N\), and the transformation between the pressure \(p(kr,\Omega)\) and ambisonics function \(A_{nm}(k)\) is approximated by,
\begin{equation}\label{eqn:ambisonics1}
    p(kr,\Omega)\approx \sum_{n=0}^{N}b_n(kr)\sum_{m=-n}^{n}A_{nm}(k)Y_{n}^{m}(\Omega),
\end{equation}
rewriting \eqref{eqn:ambisonics1} in matrix form and solving for $A_{nm}$, the linear encoding the microphone signals into ambisonics becomes,
\begin{equation}
    \mathbf{a}_{nm}=\mathbf{Y}^H\mbox{diag}(\mathbf{b}_n)^{-1}\mathbf{p}
\end{equation}where $\mathbf{Y}$ is a $Q \times (N+1)^2$ matrix of spherical harmonics, 
\begin{equation}
    {\bf Y} = \begin{bmatrix}
    {Y_0^0({\Omega_1})} & Y_1^{- 1}({\Omega_1}) & \cdots & Y_N^N({\Omega_1}) \cr Y_0^0({\Omega_2}) & Y_1^{- 1}({\Omega_2}) & \cdots & Y_N^N({\Omega_2}) \cr \vdots & \vdots & \ddots & \vdots \cr Y_0^0({\Omega_Q}) &Y_1^{- 1}({\Omega_Q}) & \cdots & Y_N^N({\Omega_Q})\end{bmatrix}
\end{equation}
$\mathbf{b}_n$ is the radial function vector, and dependency on the spatial frequency $k$ is omitted for brevity. As in this paper we consider signals to be outputs of a short-time Fourier transform, we further denote our ambisonics features as $\mathbf{A} \in \mathbb {C}^{ T \times F \times (N+1)^2}$. More specifically, for $N=1$ and a given time frame $t$ and frequency bin $f=kc/(2\pi)$ our features are contained in the following ambisonics channels:
\begin{equation}
     \boldsymbol{A}(t,f) =\begin{bmatrix}A_\text{0,0}(t,f)\\A_\text{1,-1}(t,f)\\ A_\text{1,0}(t,f)\\ A_\text{1,1}(t,f) \end{bmatrix},
\end{equation}
these correspond to the $W$ (omnidirectional) and $Y,Z,X$ (dipole) components of the first order ambisonics approximation. It is worthwhile noting that once microphone array signals have been encoded into ambisonics, no a-priori knowledge on the structure of the capturing array is needed in order to perform any downstream spatial processing. Thus, ambisonics are effectively agnostic to both recording and playback devices, making any embeddings derived from them equally generalizable.

\subsection{Ablations across model architectures}
\label{sec:model_ablations}

We ablate over using static intensity vectors in place of the learned encoder, just the learned encoder without spatial regressors, and compare them with our current architecture. Results are shown in \cref{tab:arch_ablations}.

\begin{table}[hp!]
  \scriptsize
  \caption{Comparison of semantic and spatial retrieval performance across data input ablations. mAP@10 refers to the text-to-audio mean average precision @ $10$ and audio-to-text mean average precision @ $10$. It is a summary metric capturing the model's semantic retrieval capabilities.}
  \label{tab:arch_ablations}
  \vspace*{2mm} 
  \centering
  \setlength{\tabcolsep}{1mm}
  \begin{tabular}{l ccc}
    \toprule
    & \multicolumn{3}{c}{\textsc{Spatial (AudioCaps + Clotho)}} \\
    \cmidrule{2-4}
    \textsc{Model} & \textsc{3D Local.} (\mdeg) & \textsc{Dist.} (cm) & \textsc{mAP@10} \\
    \midrule
    Static Intensity Vectors & 27.86 & 60.2 & 23.43 \\ 
    Pretrained Spatial Branch \& only CLIP Loss  & 27.4  & 54.31 & 24.93 \\ 
    Pretrained Spatial Branch \& Spatial Losses & 23.2  & 47.71 & 24.81 \\  
    \bottomrule
  \end{tabular}
\end{table}

\begin{table}[hp!]
  \scriptsize
  \caption{The retrieval metrics here demonstrate why mixing both mono and spatial audio is required to achieve the best possible retrieval performance. We see that the model needs to see both the regular and spatially augmented versions of the audio and captions to achieve best performance.}
  \label{tab:data_ablation}
  \vspace*{2mm} 
  \centering
  \setlength{\tabcolsep}{1mm} 
  \begin{tabular}{    
    >{\raggedright\arraybackslash}m{10mm}%
    >{\raggedright\arraybackslash}m{15mm}%
    >{\raggedright\arraybackslash}m{15mm}%
    m{2mm}
    >{\centering\arraybackslash}m{5mm}%
    >{\centering\arraybackslash}m{5mm}%
    >{\centering\arraybackslash}m{5mm}%
    >{\centering\arraybackslash}m{5mm}%
    >{\centering\arraybackslash}m{5mm}%
    >{\centering\arraybackslash}m{5mm}%
    m{2mm}
    >{\centering\arraybackslash}m{5mm}%
    >{\centering\arraybackslash}m{5mm}%
    >{\centering\arraybackslash}m{5mm}%
    >{\centering\arraybackslash}m{5mm}%
    >{\centering\arraybackslash}m{5mm}%
    >{\centering\arraybackslash}m{5mm}%
  }
    \toprule
    \multicolumn{4}{c}{} & \multicolumn{6}{c}{\textsc{AudioCaps}} && \multicolumn{6}{c}{\textsc{Clotho}} \\
    & \multicolumn{2}{c}{\textsc{Train Data}} && \multicolumn{3}{c}{\textsc{Text-to-Audio}} & \multicolumn{3}{c}{\textsc{Audio-to-Text}} && \multicolumn{3}{c}{\textsc{Text-to-Audio}} & \multicolumn{3}{c}{\textsc{Audio-to-Text}} \\
    \cmidrule{2-3}\cmidrule{5-10}\cmidrule{12-17}
    \textsc{Model} & \textsc{Audio} & \textsc{Captions} && R@1 & R@5 & R@10 & R@1 & R@5 & R@10 && R@1 & R@5 & R@10 & R@1 & R@5 & R@10\\
    \midrule
    CLAP  & Non-spatial & Non-spatial && 32.7 & 68.8 & 81.5 & 40.7 & 74.0 & 84.7 && 14.4 & 37.6 & 50.7 & 18.3 & 40.5 & 55.1 \\
    ELSA  & Spatial     & Non-spatial && 27.1 & 62.7 & 76.1 & 36.6 & 68.7 & 78.4 && 11.3 & 32.6 & 44.4 & 12.4 & 28.3 & 50.0 \\
    ELSA  & Spatial     & Spatial     && 25.3 & 59.3 & 72.5 & 34.8 & 64.5 & 75.2 &&  9.9 & 31.0 & 39.8 & 12.1 & 35.3 & 47.3 \\
    ELSA  & Mixed       & Mixed       && 33.2 & 68.2 & 81.0 & 40.9 & 74.4 & 86.1 && 15.0 & 36.7 & 50.8 & 20.1 & 43.2 & 55.4 \\
    \bottomrule
  \end{tabular}
\end{table}

\label{tab:mono_spatial_ablation}

\subsection{ELSA architecture}
\label{sec:elsa_architecture}

\begin{figure}[h]
\centering
\includegraphics[width=\linewidth, clip, trim={3mm 50mm 15mm 30mm}]{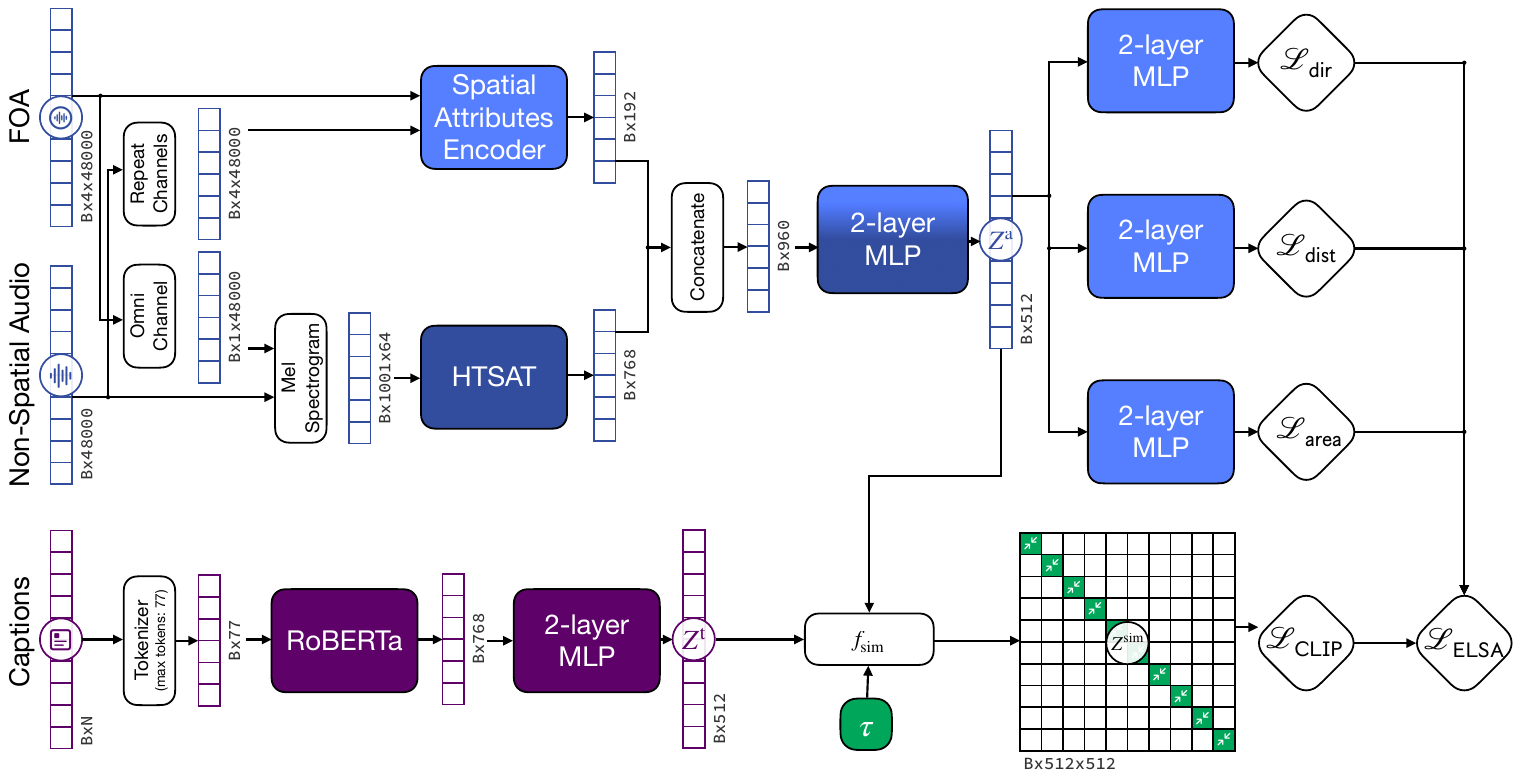}
\caption{Full architecture diagram for ELSA. Filled blocks include trainable parameters.}
\label{fig:elsa_architecture}
\end{figure}

The full architecture of the ELSA model is shown in \cref{fig:elsa_architecture}.

\subsection{Spatial attributes branch of audio encoder details}
\label{sec:spatial_attributes_branch_details}

\begin{figure}[h]
\centering
\includegraphics[width=\linewidth, clip, trim={10mm 60mm 0mm 60mm}]{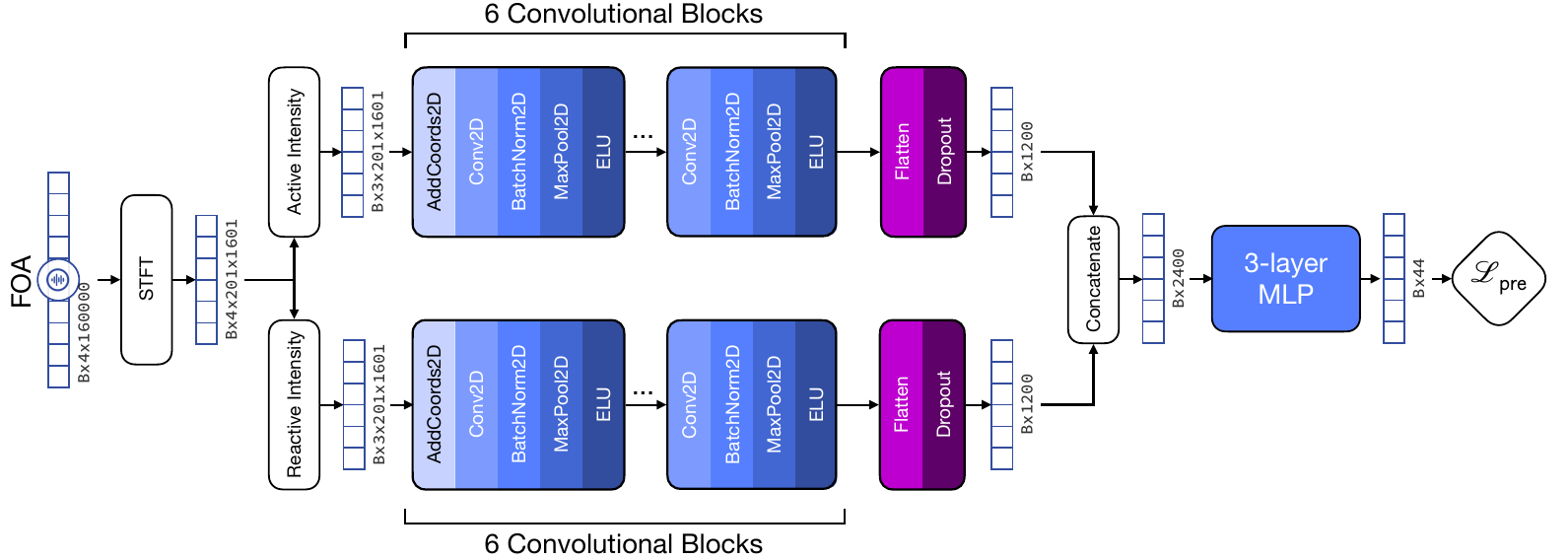}
\caption{Architecture diagram for Spatial Attributes Branch. Filled blocks include trainable parameters. The \texttt{AddCoords2D} block is described in~\cite{liu2018intriguing}.}
\label{fig:spatial_encoder_arch}
\end{figure}

The full architecture for the spatial attributes branch of the audio encoder is shown below in~\cref{fig:spatial_encoder_arch}.

The spatial attributes branch has 485,828 parameters, and was pre-trained with a learning rate of \(10^{-3}\) on the LAMB optimizer~\cite{you2020large} with weight decay factor of 0.01 and without scheduling the learning rate. The batch size was 1024 and the model was trained for 100 epochs on a single node with 8 NVIDIA V100 GPUs and 80CPUs. Training took 12h 20min. The training set was composed of 134,712 10-second segments from the first-order ambisonics samples of the Spatial LibriSpeech train set. We used a multi-task regression loss, \(\mathcal{L}_\mathrm{pre}\), to jointly learn the azimuth, elevation, distance, room volume, and 20-third octave bins (between 100Hz and 8kHz) for direct-to-reverberant ratio and T30. \(\mathcal{L}_\mathrm{pre}\) is the sum of the cosine loss for azimuth and elevation and the mean squared error for all other predictions.

\subsection{Fine-grained of direction-of-arrival error analysis}
\label{sec:doa_error_analysis}

We analyze the errors of a two-layer MLP trained to regress the direction-of-arrival (same setting as the last column in \cref{tab:spatial_audio_regression}).  We observe how the errors vary along the following dimensions: source azimuth, source elevation, source distance, room floor area, room mean T30, and TUT Sound Events 2018 semantic classes. Results are rendered as boxplots in \cref{fig:doa_boxplots} below.

\begin{figure}[hp!]
     \centering
     \begin{subfigure}[b]{0.49\textwidth}
         \centering
         \includegraphics[width=\textwidth, trim={0mm 8mm 3mm 3mm},clip]{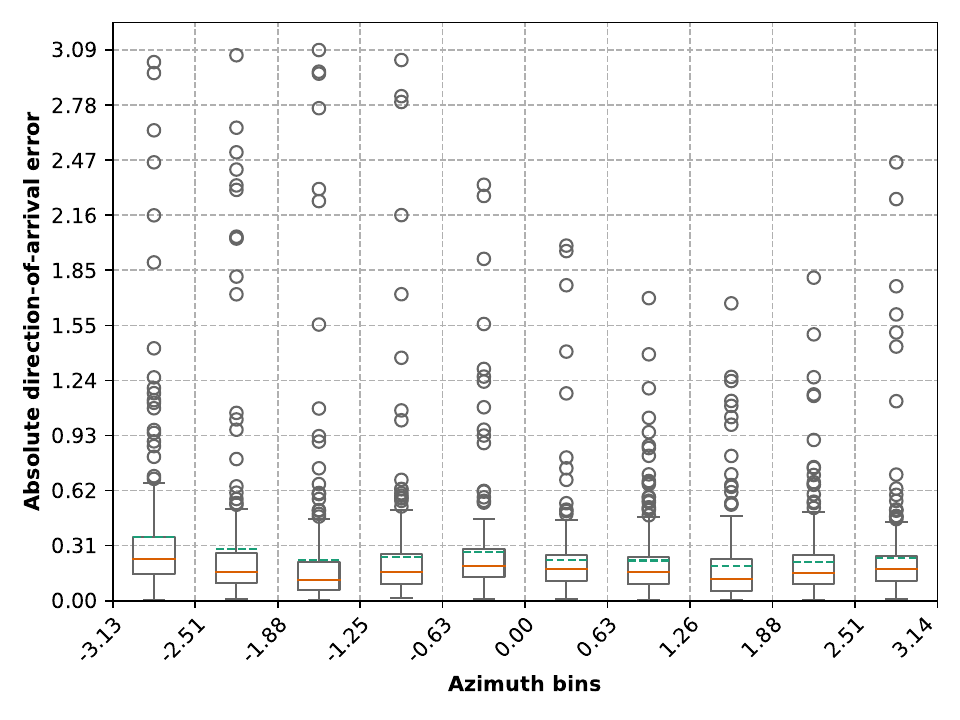}
         \caption{Azimuth (rad)}
     \end{subfigure}
     \hfill
     \begin{subfigure}[b]{0.49\textwidth}
         \centering
        \includegraphics[width=\textwidth,trim={0mm 8mm 3mm 3mm},clip]{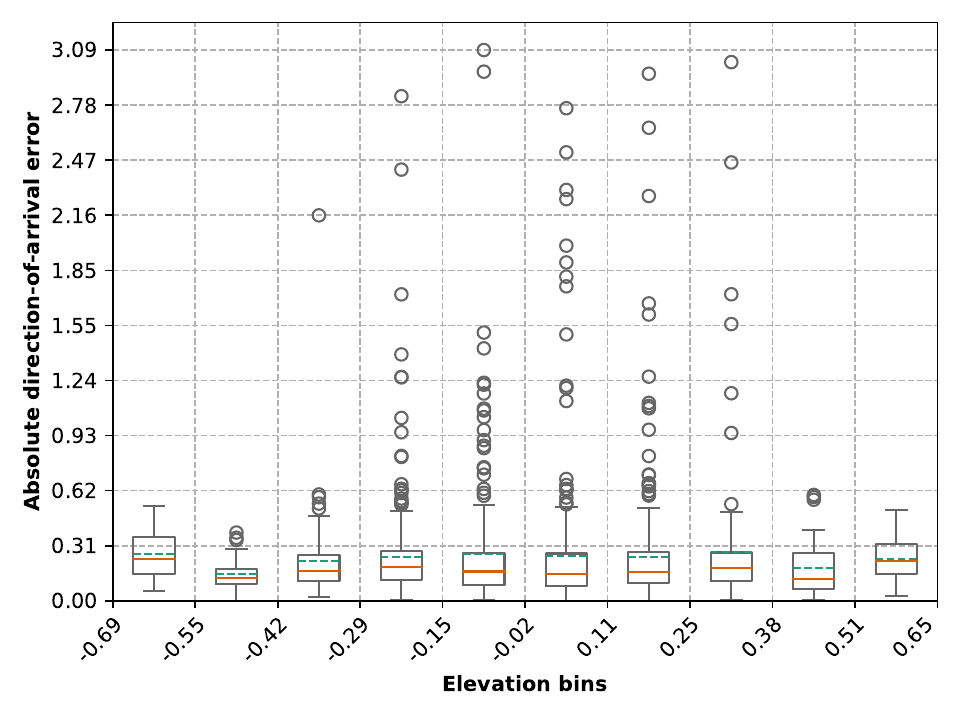}
        \caption{Elevation (rad)}
     \end{subfigure}
     \begin{subfigure}[b]{0.49\textwidth}
         \centering
         \includegraphics[width=\textwidth,trim={0mm 8mm 3mm 3mm},clip]{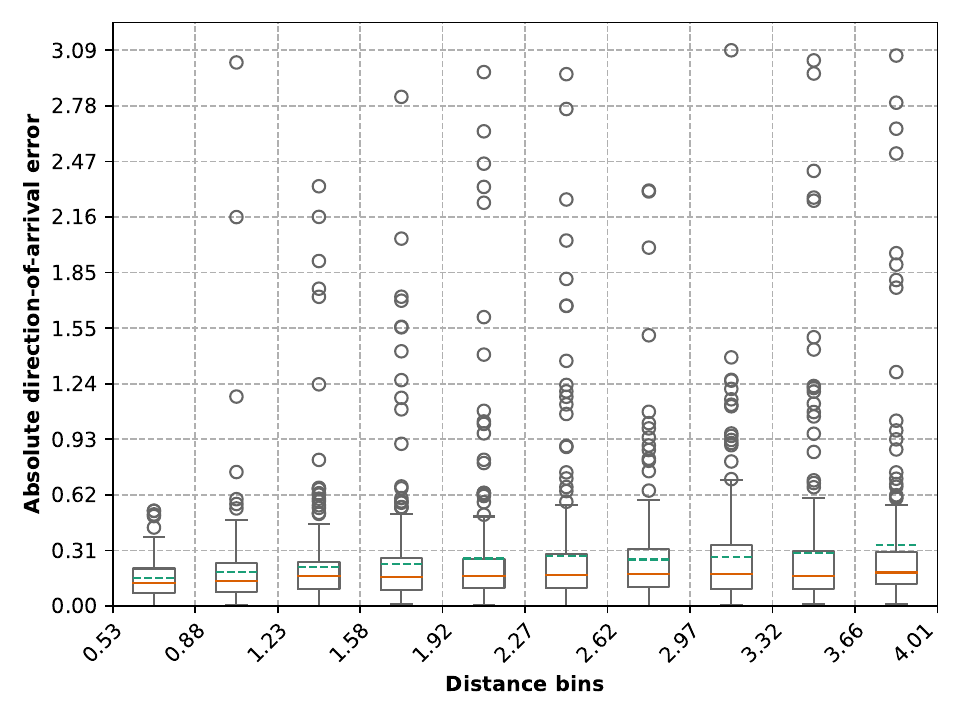}
         \caption{Distance (m)}
     \end{subfigure}
     \hfill
     \begin{subfigure}[b]{0.49\textwidth}
         \centering
         \includegraphics[width=\textwidth,trim={0mm 8mm 3mm 3mm},clip]{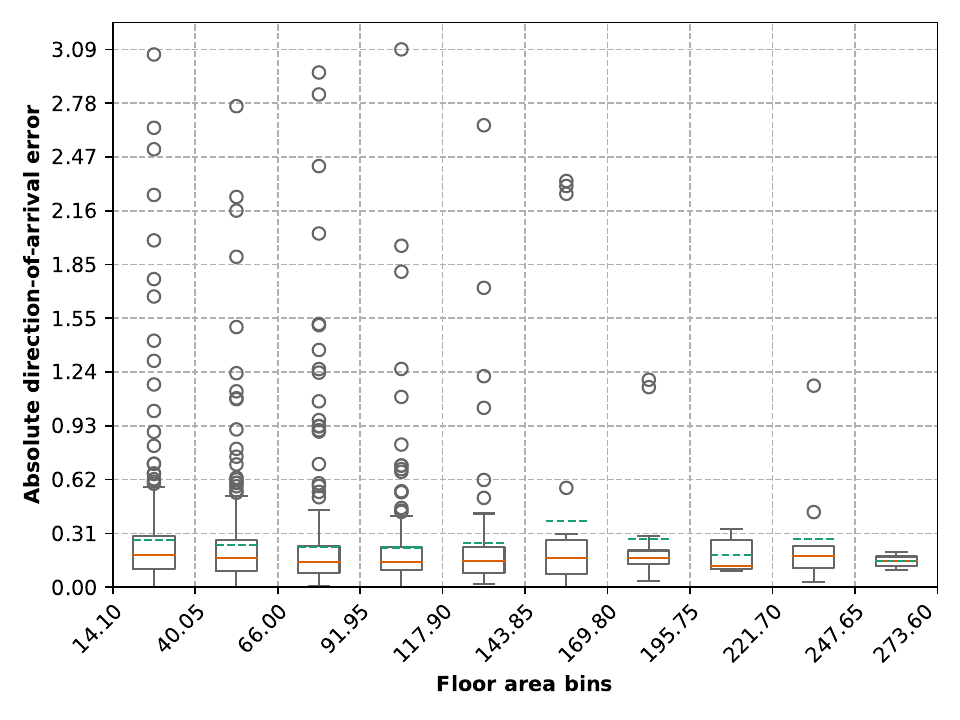}
         \caption{Floor area (m\(^\mathrm{2}\))}
     \end{subfigure}
     \begin{subfigure}[b]{0.49\textwidth}
         \centering
         \includegraphics[width=\textwidth,trim={0mm 8mm 3mm 3mm},clip]{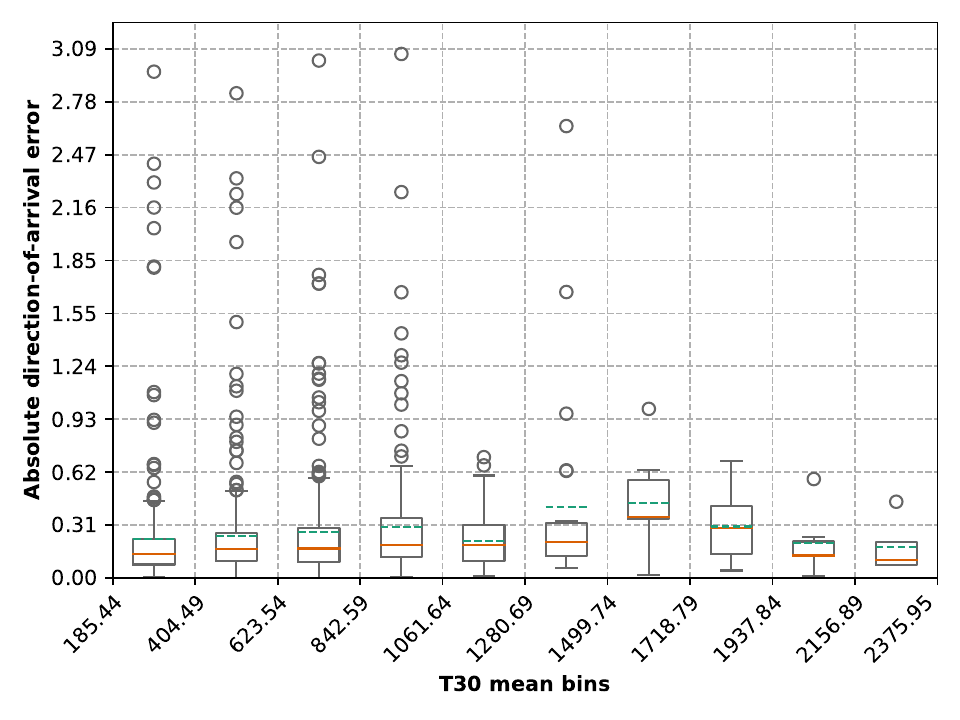}
         \caption{Mean T30 (ms)}
     \end{subfigure}
     \hfill
     \begin{subfigure}[b]{0.49\textwidth}
         \centering
         \includegraphics[width=\textwidth,trim={0mm 8mm 3mm 3mm},clip]{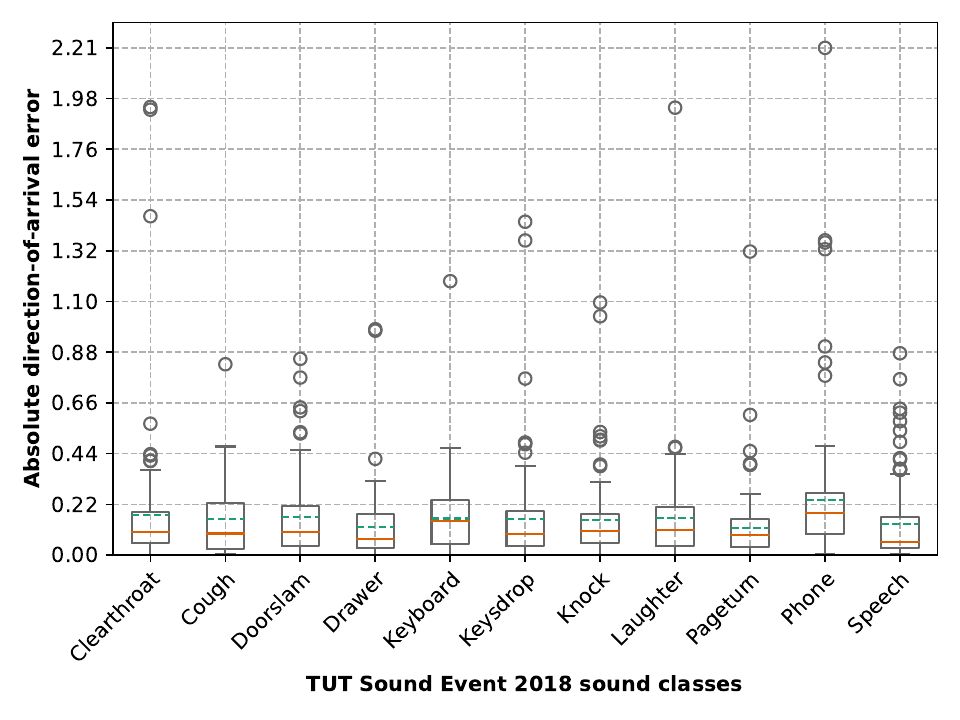}
         \caption{TUT Sound Events 2018 Semantic classes}
     \end{subfigure}
     \caption{Boxplots of absolute direction-of arrival errors predicted by 2-layer MLP. Figs. (a)--(e) show the Spatial Audiocaps and Spatial Clotho test sets errors by different categories. Fig. (f) shows the predictions of the test set of TUT Sounds 2018 by different semantic classes. For all figures, boxes represent the interquartile range, solid orange lines are the median, and dashed green lines are the mean.}
     \label{fig:doa_boxplots}
\end{figure}

\cref{tab:doa_finegrained_errors} further report mean, standard deviation, and number of samples per bin about the direction-of-arrival errors across the previously analysed dimensions.

\begin{table}[hp!]
  \scriptsize
  \centering
  \caption{Mean and standard deviation of absolute direction-of arrival errors (in radians) predicted by 2-layer MLP. Tables (a)--(e) show the Spatial Audiocaps and Spatial Clotho test sets errors by different dimensions. Table (f) shows the predictions of the test set of TUT Sounds 2018 by different semantic classes.}
  \label{tab:doa_finegrained_errors}
  
  \begin{subtable}[b]{0.49\textwidth}
    \centering
    \caption{DOA error by azimuth}
    \begin{tabular}{
        >{\centering\arraybackslash}m{19mm}%
        >{\centering\arraybackslash}m{11mm}%
        >{\centering\arraybackslash}m{11mm}%
        >{\centering\arraybackslash}m{10mm}%
    }
        \toprule
        \textsc{Azimuth Ranges} (rad) & \textsc{DOA Err. Mean} & \textsc{DOA Err. Std. dev} & \textsc{\# Bin Samples}\\
        \midrule
		\([-3.13, -2.51)\) & 0.36 & 0.44 & 272 \\
		\([-2.51, -1.88)\) & 0.29 & 0.46 & 251 \\
		\([-1.88, -1.25)\) & 0.23 & 0.43 & 258 \\
		\([-1.25, -0.63)\) & 0.24 & 0.37 & 245 \\
		\([-0.63, +0.00)\) & 0.27 & 0.31 & 215 \\
		\([+0.00, +0.63)\) & 0.23 & 0.25 & 224 \\
		\([+0.63, +1.26)\) & 0.23 & 0.24 & 270 \\
		\([+1.26, +1.88)\) & 0.19 & 0.23 & 220 \\
		\([+1.88, +2.51)\) & 0.22 & 0.22 & 259 \\
		\([+2.51, +3.14)\) & 0.24 & 0.29 & 244 \\
        \bottomrule
    \end{tabular}
  \end{subtable}
  \hfill
  \begin{subtable}[b]{0.49\textwidth}
    \centering
    \caption{DOA error by elevation}
    \begin{tabular}{
        >{\centering\arraybackslash}m{19mm}%
        >{\centering\arraybackslash}m{11mm}%
        >{\centering\arraybackslash}m{11mm}%
        >{\centering\arraybackslash}m{10mm}%
    }
        \toprule
        \textsc{Elevation Ranges} (rad) & \textsc{DOA Err. Mean} & \textsc{DOA Err. Std. dev} & \textsc{\# Bin Samples}\\
        \midrule
		\([-0.69, -0.55)\) & 0.26 & 0.15 & 12 \\
		\([-0.55, -0.42)\) & 0.15 & 0.11 & 20 \\
		\([-0.42, -0.29)\) & 0.22 & 0.25 & 88 \\
		\([-0.29, -0.15)\) & 0.24 & 0.28 & 312 \\
		\([-0.15, -0.02)\) & 0.26 & 0.35 & 660 \\
		\([-0.02, +0.11)\) & 0.25 & 0.38 & 735 \\
		\([+0.11, +0.25)\) & 0.24 & 0.33 & 456 \\
		\([+0.25, +0.38)\) & 0.28 & 0.40 & 126 \\
		\([+0.38, +0.51)\) & 0.18 & 0.16 & 35 \\
		\([+0.51, +0.65)\) & 0.24 & 0.13 & 14 \\
        \bottomrule
    \end{tabular}
  \end{subtable}

  \begin{subtable}[b]{0.49\textwidth}
    \centering
    \caption{DOA error by distance}
    \begin{tabular}{
        >{\centering\arraybackslash}m{19mm}%
        >{\centering\arraybackslash}m{11mm}%
        >{\centering\arraybackslash}m{11mm}%
        >{\centering\arraybackslash}m{10mm}%
    }
        \toprule
        \textsc{Distance Ranges} (m) & \textsc{DOA Err. Mean} & \textsc{DOA Err. Std. dev} & \textsc{\# Bin Samples}\\
        \midrule
		\([0.53, 0.88)\) & 0.16 & 0.12 & 100 \\
		\([0.88, 1.23)\) & 0.19 & 0.26 & 252 \\
		\([1.23, 1.58)\) & 0.21 & 0.26 & 340 \\
		\([1.58, 1.92)\) & 0.23 & 0.30 & 322 \\
		\([1.92, 2.27)\) & 0.26 & 0.39 & 314 \\
		\([2.27, 2.62)\) & 0.28 & 0.38 & 287 \\
		\([2.62, 2.97)\) & 0.26 & 0.29 & 268 \\
		\([2.97, 3.32)\) & 0.27 & 0.33 & 208 \\
		\([3.32, 3.66)\) & 0.30 & 0.45 & 203 \\
		\([3.66, 4.01)\) & 0.34 & 0.51 & 164 \\
        \bottomrule
    \end{tabular}
  \end{subtable}
  \hfill
  \begin{subtable}[b]{0.49\textwidth}
    \centering
    \caption{DOA error by room floor area}
    \begin{tabular}{
        >{\centering\arraybackslash}m{19mm}%
        >{\centering\arraybackslash}m{11mm}%
        >{\centering\arraybackslash}m{11mm}%
        >{\centering\arraybackslash}m{10mm}%
    }
        \toprule
        \textsc{Floor Area Ranges} (\sqm) & \textsc{DOA Err. Mean} & \textsc{DOA Err. Std. dev} & \textsc{\# Bin Samples}\\
        \midrule
		\([14.10, 40.05)\) & 0.27 & 0.36 & 820 \\
		\([40.05, 66.00)\) & 0.24 & 0.31 & 983 \\
		\([66.00, 91.95)\) & 0.23 & 0.33 & 346 \\
		\([91.95, 117.90)\) & 0.23 & 0.33 & 173 \\
		\([117.90, 143.85)\) & 0.25 & 0.40 & 69 \\
		\([143.85, 169.80)\) & 0.38 & 0.65 & 30 \\
		\([169.80, 195.75)\) & 0.28 & 0.32 & 18 \\
		\([195.75, 221.70)\) & 0.19 & 0.10 & 7 \\
		\([221.70, 247.65)\) & 0.28 & 0.31 & 10 \\
		\([247.65, 273.60)\) & 0.15 & 0.05 & 2 \\
        \bottomrule
    \end{tabular}
  \end{subtable}

  \begin{subtable}[b]{0.49\textwidth}
    \centering
    \caption{DOA error by T30}
    \begin{tabular}{
        >{\centering\arraybackslash}m{19mm}%
        >{\centering\arraybackslash}m{11mm}%
        >{\centering\arraybackslash}m{11mm}%
        >{\centering\arraybackslash}m{10mm}%
    }
        \toprule
        \textsc{T30 Ranges} (ms) & \textsc{DOA Err. Mean} & \textsc{DOA Err. Std. dev} & \textsc{\# Bin Samples}\\
        \midrule
		\([185.44, 404.49)\) & 0.23 & 0.35 & 585 \\
		\([404.49, 623.54)\) & 0.25 & 0.34 & 1030 \\
		\([623.54, 842.59)\) & 0.27 & 0.35 & 521 \\
		\([842.59, 1061.64)\) & 0.29 & 0.35 & 191 \\
		\([1061.64, 1280.69)\) & 0.22 & 0.15 & 78 \\
		\([1280.69, 1499.74)\) & 0.41 & 0.57 & 25 \\
		\([1499.74, 1718.79)\) & 0.44 & 0.26 & 9 \\
		\([1718.79, 1937.84)\) & 0.30 & 0.19 & 9 \\
		\([1937.84, 2156.89)\) & 0.20 & 0.18 & 6 \\
		\([2156.89, 2375.95)\) & 0.18 & 0.15 & 4 \\
        \bottomrule
    \end{tabular}
  \end{subtable}
  \hfill
  \begin{subtable}[b]{0.49\textwidth}
    \centering
    \caption{DOA error by TUT Sound Events 2018 semantic class}
    \begin{tabular}{
        >{\centering\arraybackslash}m{19mm}%
        >{\centering\arraybackslash}m{11mm}%
        >{\centering\arraybackslash}m{11mm}%
        >{\centering\arraybackslash}m{10mm}%
    }
        \toprule
        \textsc{Semantic class} & \textsc{DOA Err. Mean} & \textsc{DOA Err. Std. dev} & \textsc{\# Bin Samples}\\
        \midrule
        Drawer &  0.12 & 0.16 & 97\\
        Laughter & 0.16 & 0.22 & 95\\
        Cough & 0.16 & 0.16 & 87\\
        Clearthroat & 0.17 & 0.29 & 115\\
        Keyboard & 0.16 & 0.15 & 97 \\
        Speech & 0.14 & 0.17 & 105 \\
        Phone & 0.24 & 0.30 & 117 \\
        Pageturn & 0.12 & 0.15 & 115 \\
        Knock & 0.15 & 0.17 & 112 \\
        Doorslam & 0.17 & 0.17 & 101 \\
        Keysdrop &  0.15 & 0.21 & 111 \\
        \bottomrule
    \end{tabular}
  \end{subtable}
\end{table}

\subsection{Spatial attributes retrieval for LAION-CLAP}
\label{sec:spatial_attributes_clap}
\cref{sec:spatial_attributes_results} showed that ELSA embeddings can be classified in a zero-shot fashion by using a templated probe caption. \cref{tab:full_spatial_attributes_retrieval} shows the same experiment applied to LAION-CLAP (with the ELSA results retained for context) by feeding LAION-CLAP the omni-channel of the spatial datasets. As expected the performance of LAION-CLAP in this setting is close to random, for instance LAION-CLAP achieves 48\% accuracy on the two class distance classification task and 28.2\% on the four-class direction classification task.

\begin{table}[hp!]
  \scriptsize
  \centering
  \caption{Complete version of \cref{tab:spatial_attributes_retrieval} with LAION-CLAP results. LAION-CLAP results are obtained by passing omni channel of spatial dataset through pre-trained model.}
  \label{tab:full_spatial_attributes_retrieval}
  \vspace*{2mm} 
  \begin{tabular}{l ccc c ccc}
     \toprule
     & \multicolumn{3}{c}{ELSA} && \multicolumn{3}{c}{LAION-CLAP} \\
     \cmidrule{2-4}\cmidrule{6-8}
     \textsc{Task} & \textsc{S-Clotho} & \textsc{S-AC} & \textsc{S-RWD} && \textsc{S-Clotho} & \textsc{S-AC} & \textsc{S-RWD} \\
     \midrule
     Distance (2-class)       &  96.0\%  &  92.9\% & 67.1\% && 48.0\% & 54.3\% & 53.0\% \\
     Direction (4-class)      &  92.0\%  &  92.8\% & 35.8\% && 28.2\% & 29.3\% & 27.3\% \\
     Elevation (2-class)      & 100.0\%  & 100.0\% & 72.1\% && 56.7\% & 51.3\% & 59.4\% \\
     Room area (2-class)      &  76.6\%  &  74.7\% & N/A    && 46.3\% & 66.5\% & N/A    \\
     Reverberation (2-class)  & 100.0\%  &  83.3\% & N/A    && 57.3\% & 52.5\% & N/A    \\
     \bottomrule
  \end{tabular}
\end{table}

\subsection{Text corpus used when testing ELSA's implicitly learned spatial attributes}
\label{implicit_texts}

As mentioned in main text, we validate that ELSA's audio embeddings can capture implicit spatial attributes, which are latent in the text-encoder. We use LLaMA-13B~\cite{touvron2023llama} to generate descriptions of 50 sounds that typically come from each of above and below. The sentences do not necessarily include explicit spatial descriptions (e.g., "A fire alarm going off", a sound typically from above but not explicitly stated). The original prompt and generated sentences can be found below. We then train a two-class classifier (top vs bottom) using the spatial audio in the train sets of our spatial-augmented datasets, and classify each of the sentences. Results show an above-random classification accuracy of 68.75\% for \emph{above} sentences, 58.82\% for \emph{below} sentences. This experiment shows that, to a degree, the text-encoder (RoBERTa) is able to leverage its semantic bias of placement of objects in the real world and encode it as spatial features that the spatial encoder understands. This task is made quite hard by the observation that our simulated dataset does not reflect the natural world (an aeroplane sound could be simulated from below) and the fact that \texttt{RoBERTa-base} has significantly fewer parameters and a smaller training set more recent large-language models such as LLaMA-13B. This experiment is purely qualitative, and we leave a more in-depth exploration of knowledge sharing between the pretrained spatial and text encoders to future work.

\begin{mdframed}[style=prompt, frametitlebackgroundcolor=prompttitlecolor, frametitle={Sentences for sounds from above}]
\scriptsize
\begin{itemize}[leftmargin=3mm]
\setlength\itemsep{0mm}
\item Sound from the upper part of a room.
\item The sound of an aeroplane flying.
\item The sound of a bird chirping.
\item The sound of a helicopter in the sky.
\item A low rumble of distant thunder.
\item A sharp crack of lightning striking.
\item The pitter-patter of rain on a roof.
\item The rhythmic whoosh of wind through trees.
\item A distant siren wailing high in the air.
\item The muffled thud of footsteps on a floor upstairs.
\item A child's laughter echoing from an upstairs room.
\item A playful meow or bark from a pet overhead.
\item The frantic buzzing of a trapped fly.
\item The soft hum of a ceiling fan rotating.
\item The rhythmic beeping of a smoke alarm.
\item The distant hooting of an owl.
\item The rustling of leaves as a squirrel scurries across a roof.
\item The muffled boom of fireworks exploding high above.
\item The melodic ringing of church bells.
\item A fighter jet screaming across the heavens.
\item The rhythmic thump of a basketball bouncing overhead.
\item The scraping of furniture being moved on a floor above.
\item The high-pitched whine of a mosquito circling your head.
\item The rhythmic drumming of raindrops on a skylight.
\item A loud thump followed by a startled yelp (someone tripped upstairs).
\item The rhythmic tap-tap-tap of a woodpecker on a tree trunk.
\item The frantic buzzing of a swarm of bees overhead.
\item The mournful cry of a seagull circling above the beach.
\item The rhythmic clatter of hail bouncing off a roof.
\item The melodic singing of a bird outside your window.
\item The gentle whoosh of a hot air balloon floating overhead.
\item The rhythmic thump-thump-thump of helicopter blades.
\item Someone walking on the roof.
\item The soft cooing of pigeons perched on a building ledge.
\item The gentle pitter-patter of rain on a tent roof.
\item The high-pitched shriek of a child on a roller coaster.
\item The rhythmic chirping of crickets in a field at night.
\item The muffled thump of a heavy object being dropped from above.
\item The rhythmic whir of a helicopter hovering nearby.
\item The faint melody of a lullaby drifting down a stairwell.
\item A sudden gust of wind whistling through the trees.
\item The rhythmic clicking of a computer mouse from upstairs.
\item The melodic ringing of a wind chime swaying in the breeze.
\item The rhythmic pounding of rain on a metal roof.
\item The rhythmic chirping of birds waking you up at dawn.
\item The muffled conversation of people walking on a floor above.
\item The muffled snoring of someone sleeping upstairs.
\item The chirping and squawking of a flock of birds taking flight.
\item The rhythmic click-clack of tap shoes dancing on a floor above.
\item The rhythmic tapping of a woodpecker searching for insects.
\end{itemize}
\end{mdframed}
\vspace*{5mm}

\begin{mdframed}[style=prompt, frametitlebackgroundcolor=prompttitlecolor, frametitle={Sentences for sounds from below}]
\scriptsize
\begin{itemize}[leftmargin=3mm]
\setlength\itemsep{0mm}
\item The sound of an underground subway train.
\item The sound of feet walking on a pavement.
\item The muffled boom of a distant explosion.
\item The rhythmic dripping of a leaky pipe in the basement.
\item The gurgling of water in a drainpipe.
\item The muffled thump of something heavy being dropped downstairs.
\item The low hum of a refrigerator running.
\item The rhythmic squeak of floorboards underfoot.
\item The muffled clinking of glasses from below.
\item The faint laughter echoing from downstairs.
\item The rhythmic drumming of a clothes dryer in the basement.
\item The rhythmic beeping of a malfunctioning appliance in the basement.
\item The scratching sound of a pet exploring under furniture.
\item The scurrying of mice in the floorboards.
\item The rhythmic dripping of a bath faucet you forgot to turn off completely.
\item The muffled roar of a furnace kicking on.
\item The rhythmic tick-tock of a grandfather clock.
\item The faint vibration of a subwoofer from a downstairs stereo.
\item The muffled chatter of people in a room below.
\item The rhythmic click-clack of a keyboard from below.
\item The muffled thump of a door closing downstairs.
\item The rhythmic pinging of a pinball machine in a basement arcade.
\item The dripping sound of melting ice from a refrigerator.
\item The rhythmic whoosh of a basement exhaust fan.
\item The faint hum of electrical wiring in the walls.
\item The rhythmic drumming of rain on a basement window.
\item The low rumble of distant traffic filtering through the floorboards.
\item The faint creaking of an old house settling.
\item The rhythmic thump of a bouncing ball from downstairs.
\item The muffled clinking of coins in a jar breaking on the floor.
\item The rhythmic whoosh of a vacuum cleaner downstairs.
\item The faint strains of music seeping up from a basement party.
\item The rhythmic beeping of a smoke alarm in the basement (hopefully a false alarm).
\item The muffled shouts of children playing downstairs.
\item The rhythmic tapping of a sewer pipe being repaired.
\item The rhythmic dripping of condensation on a cold water pipe.
\item The rhythmic click of a deadbolt lock being secured downstairs.
\item The muffled roar of a lawnmower outside.
\item The rhythmic scratching of a pet trying to dig a hole in the carpet.
\item The muffled clinking of silverware being dropped in the sink.
\item The rhythmic thump of a basketball bouncing on the floor below.
\item The rhythmic ping-pong of a table tennis match in progress downstairs.
\item The rhythmic whirring of a washing machine agitating clothes downstairs.
\item The muffled snoring of someone sleeping downstairs.
\item The rhythmic drumming of rain on a basement windowpane at night.
\item The faint, high-pitched whine of a mosquito buzzing around your ankles.
\item The muffled clatter of pots and pans being moved around in the kitchen.
\item The rhythmic whoosh of a sprinkler system watering the lawn.
\end{itemize}
\end{mdframed}

\subsection{Semantic retrieval for spatial data}
\label{sec:semantic_retrieval_appendix}

We report ELSA's semantic retrieval on Spatial-Clotho, and Spatial-Audiocaps in \cref{tab:semantic_retrieval_spatial}. Likewise, \cref{tab:semantic_retrieval_spatial_rwd} shows the retrieval scores of ELSA on our spatial real-world dataset.

\begin{table}[hp!]
  \scriptsize
  \caption{Semantic Retrieval Metrics calculated over spatially augmented version of Clotho and AudioCaps eval sets -- identical in size as the non-spatial sets.}
  \label{tab:semantic_retrieval_spatial}
  \vspace*{2mm} 
  \centering
  \setlength{\tabcolsep}{1mm} 
  \begin{tabular}{    
    >{\raggedright\arraybackslash}m{10mm}%
    >{\raggedright\arraybackslash}m{25mm}%
    >{\centering\arraybackslash}m{5mm}%
    >{\centering\arraybackslash}m{5mm}%
    >{\centering\arraybackslash}m{5mm}%
    >{\centering\arraybackslash}m{5mm}%
    >{\centering\arraybackslash}m{5mm}%
    >{\centering\arraybackslash}m{5mm}%
    m{2mm}
    >{\centering\arraybackslash}m{5mm}%
    >{\centering\arraybackslash}m{5mm}%
    >{\centering\arraybackslash}m{5mm}%
    >{\centering\arraybackslash}m{5mm}%
    >{\centering\arraybackslash}m{5mm}%
    >{\centering\arraybackslash}m{5mm}%
    m{2mm}
  }
    \toprule
    \multicolumn{2}{c}{} & \multicolumn{6}{c}{\textsc{Spatial-AudioCaps}} && \multicolumn{6}{c}{\textsc{Spatial-Clotho}}           \\
    \multicolumn{2}{c}{} & \multicolumn{3}{c}{\textsc{Text-to-Audio}} & \multicolumn{3}{c}{\textsc{Audio-to-Text}} && \multicolumn{3}{c}{\textsc{Text-to-Audio}} & \multicolumn{3}{c}{\textsc{Audio-to-Text}} \\
    \cmidrule{3-8}\cmidrule{10-15}
    \textsc{Model} & \textsc{Train Data} & R@1 & R@5 & R@10 & R@1 & R@5 & R@10 && R@1 & R@5 & R@10 & R@1 & R@5 & R@10\\
    \midrule
    ELSA & Sp(Clotho + AC) & 25.3  & 58.9  & 73.5  & 32.6 & 61.5  & 73.0 && 9.4 & 25.9 & 38.0 & 10.5 & 27.8 & 40.2 \\ 
    ELSA & Sp(Clotho + AC + FS) & 24.2 & 56.7 & 71.82 & 30.5  & 59.42  & 72.8  && 11.26 & 30.72 & 43.07 & 12.6 & 32.0 & 44.08 \\ 
    \bottomrule
  \end{tabular}
\end{table}

\begin{table}[hp!]
  \scriptsize
  \caption{Semantic Retrieval Metrics calculated over our spatial real-world dataset.}
  \label{tab:semantic_retrieval_spatial_rwd}
  \vspace*{2mm} 
  \hspace*{-15mm}
  \centering
  \setlength{\tabcolsep}{1mm} 
  \begin{tabular}{    
    >{\raggedright\arraybackslash}m{10mm}%
    >{\raggedright\arraybackslash}m{25mm}%
    >{\centering\arraybackslash}m{5mm}%
    >{\centering\arraybackslash}m{5mm}%
    >{\centering\arraybackslash}m{5mm}%
    >{\centering\arraybackslash}m{5mm}%
    >{\centering\arraybackslash}m{5mm}%
    >{\centering\arraybackslash}m{5mm}%
  }
    \toprule
    \multicolumn{2}{c}{} & \multicolumn{6}{c}{\textsc{Spatial-RWD}}              \\
    \multicolumn{2}{c}{} & \multicolumn{3}{c}{\textsc{Text-to-Audio}} & \multicolumn{3}{c}{\textsc{Audio-to-Text}} \\
    \cmidrule{3-8}
    \textsc{Model} & \textsc{Train Data} & R@1 & R@5 & R@10 & R@1 & R@5 & R@10 \\
    \midrule
    ELSA & Sp(Clotho + AC) & 18.6 & 54.3 & 75.7 & 25.7 & 55.7 & 71.4 \\ 
    ELSA & Sp(Clotho + AC + FS) & 41.4 & 68.6 & 88.6 & 35.7 & 67.14 & 80.0 \\ 
    \bottomrule
  \end{tabular}
\end{table}

\subsection{Further analysis on embedding clusters}
\label{sec:embedding_clusters}

\begin{figure}[t!]
\centering
\includegraphics[width=0.6\textwidth, trim={3mm 26mm 3mm 26mm},clip]{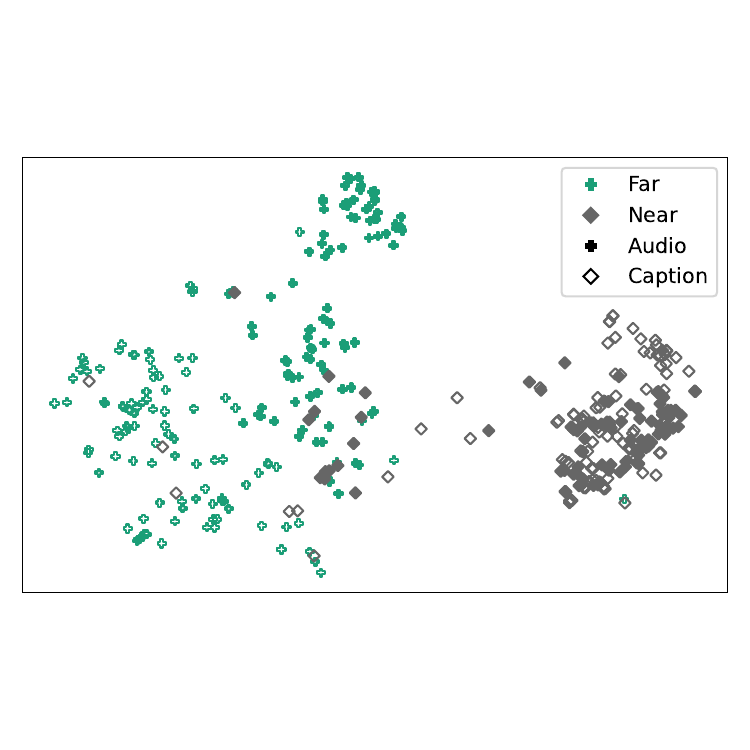}
\caption{UMAP projection of ELSA embeddings of the test splits of Spatial-Clotho and Spatial-AudioCaps. Filled markers are obtained from spatial audio, and hollow markers are obtained from spatial captions. The UMAP projection was fitted with the train splits of Spatial-Clotho and Spatial-Audio caps, and we made use of supervised dimension reduction to highlight the \textit{distance} differences rather than the semantic differences in the embeddings.}
\label{fig:distance_clusters}
\end{figure}

In \cref{sec:embedding_structure} we analysed the UMAP projection of the ELSA embeddings of the test set of Spatial AudioCaps and Spatial Clotho. \cref{tab:cluster_wasserstein_distances}~(a) shows the Wasserstein distances computed directly in the 512-dimensional space, where we see the data clusters by direction with lower Wasserstein distances between ``front'' and ``back''. Similarly, \cref{fig:distance_clusters} and \cref{tab:cluster_wasserstein_distances}~(b) show the ELSA embeddings can be clustered according to spatial distance characteristics.

\begin{table}[hp!]
  \scriptsize
  \centering
  \caption{Wasserstein distances of 512-dimensional ELSA embeddings, clustered by either (a) direction or (b) distance.}
  \label{tab:cluster_wasserstein_distances}
  \hfill
  \begin{subtable}[t]{0.4\textwidth}
    \centering
    \caption{Direction clustering distances}
    \begin{tabular}{ccccc}
        \toprule
        & \textsc{Left} & \textsc{Right} & \textsc{Front} & \textsc{Back}\\
        \midrule
		\textsc{Left} &  0.00 & 1.04 & 0.94 & 0.98 \\
		\textsc{Right} &  1.04 & 0.00 & 0.92 & 0.97 \\
		\textsc{Front} & 0.94 & 0.92 & 0.00 & 0.81 \\
		\textsc{Back} & 0.98 & 0.97 & 0.81 & 0.00 \\
        \bottomrule
    \end{tabular}
  \end{subtable}
  \hfill
  \begin{subtable}[t]{0.4\textwidth}
    \centering
    \caption{Distance clustering distances}
    \begin{tabular}{ccc}
        \toprule
        & \textsc{Near} & \textsc{Far}\\
        \midrule
		\textsc{Near} &  0.00 & 1.10  \\
		\textsc{Far} & 1.10 & 0.00 \\
        \bottomrule
    \end{tabular}
  \end{subtable}
  \hfill
\end{table}

\newpage
\subsection{Swapping of Spatial Direction Experiments}
\label{sec:direction_transposition}

As already introduced in \cref{sec:embedding_structure}. Our spatial direction swapping pipeline consists of the following steps:

\begin{enumerate}
    \item We obtain the ELSA embeddings four directions (``left'', ``right'', ``front'', ``back'') with the template: ``\texttt{A sound coming from the \textit{direction}}''. These are our direction prototypes.
    \item Train a 4-class direction classifier with a 2-layer MLP (33k parameters) on the training set of Spatial-AudioCaps and Spatial-Clotho. We obtained a 96.7\% classification accuracy of the test set of Spatial-AudioCaps and Spatial-Clotho.
    \item For every correctly-classified sample in the test sets, we obtain their ELSA embedding, subtract the prototype of the original direction, and add a prototype for the new direction.
\end{enumerate}

Additionally, we measure any changes in sound \emph{semantics} by computing the difference in recall@10 between the ELSA audio embedding and the ELSA embedding of the sample's non-spatial description.

Detailed results are shown in table \cref{tab:direction_change}. Overall, we find an average 99.7\% of the samples are correctly classified with the new direction after transposition, and an average change of -0.2\% in recall@10. These results show ELSA directional attributes can be linearly swapped without affecting the semantics of sound.

\begin{table}[ht!]
\scriptsize
\centering
\caption{Direction swapping of ELSA embeddings. See \cref{sec:direction_transposition} for a detailed explanation of how we swapped the embedding directions. $\notin$ is the number of test samples misclassified by our direction classifier, and subsequently excluded. N is the number of samples that were used for direction transposition. R@10 is the recall@10 computed over the corresponding non-spatial captions. $\theta$ is the classification accuracy of the transposed sample. $\Delta_{R@10}$ is the change in recall@10 after performing the change of direction.}
\label{tab:direction_change}
\vspace*{2mm}
\begin{tabular}{    
    m{2mm} 
    >{\raggedright\arraybackslash}m{6mm} 
    >{\centering\arraybackslash}m{4mm} 
    >{\centering\arraybackslash}m{4mm} 
    >{\centering\arraybackslash}m{4mm} 
    m{0mm}
    >{\centering\arraybackslash}m{6mm}%
    >{\centering\arraybackslash}m{6mm}%
    m{0mm}
    >{\centering\arraybackslash}m{6mm}%
    >{\centering\arraybackslash}m{6mm}%
    m{0mm}
    >{\centering\arraybackslash}m{6mm}%
    >{\centering\arraybackslash}m{6mm}%
    m{0mm}
    >{\centering\arraybackslash}m{6mm}%
    >{\centering\arraybackslash}m{6mm}%
}
\toprule
\multicolumn{6}{c}{} & \multicolumn{11}{c}{\textsc{New Direction}}\\
\cmidrule{7-17}
\multicolumn{5}{c}{} && \multicolumn{2}{c}{\textsc{Left}} && \multicolumn{2}{c}{\textsc{Front}} && \multicolumn{2}{c}{\textsc{Right}} && \multicolumn{2}{c}{\textsc{Back}} \\
\cmidrule{7-8}\cmidrule{10-11}\cmidrule{13-14}\cmidrule{16-17}
&& $\notin$ & N & R@10 && $\theta$ & $\Delta_{R@10}$ && $\theta$ & $\Delta_{R@10}$ && $\theta$ & $\Delta_{R@10}$ && $\theta$ & $\Delta_{R@10}$ \\
\midrule
\parbox[m]{2mm}{\multirow{4}{*}{\rotatebox[origin=c]{90}{\textsc{Original Dir.}}}} & \textsc{Left} & 14 & 156 & 94.9\% && -- & -- && 100.0\% & +1.3\% && 100.0\% & +0.0\% && 100.0\% & -1.3\%  \\[2mm]
& \textsc{Front} & 13 & 486 & 81.1\% && 100.0\% & +1.9\% && -- & -- && 100.0\% & +0.6\% && 100.0\% & -0.4\% \\[2mm]
& \textsc{Right} & 12 & 196 & 92.3\% && 100.0\% & -0.5\% && 100.0\% & +0.5\% && -- & -- && 100.0\% & -0.5\% \\[2mm]
& \textsc{Back} & 9 & 564 & 81.6\% && 97.3\% & -1.4\% && 100.0\% & -0.5\% && 100.0\% & -0.9\% && -- & --  \\
\bottomrule
\end{tabular}
\end{table}

Furthermore, we wanted to verify what happened if we removed the original direction but did not add back a new direction. ~\Cref{tab:direction_removal} shows this ablation. Interestingly, the classification does not result in random classification accuracy but rather 0\% accuracy for all four original directions. 

\begin{table}[htb]
\scriptsize
\centering
\caption{Direction removal of ELSA embeddings. See \cref{sec:direction_transposition} for a detailed explanation of how we swapped the embedding directions. $\notin$ is the number of test samples misclassified by our direction classifier, and subsequently excluded. N is the number of samples that were used for direction transposition. R@10 is the recall@10 computed over the corresponding non-spatial captions. $\theta$ is the classification accuracy of the transposed sample. $\Delta_{R@10}$ is the change in recall@10 after performing the change of direction.}
\label{tab:direction_removal}
\begin{tabular}{    
    >{\raggedright\arraybackslash}m{6mm} 
    >{\centering\arraybackslash}m{4mm} 
    >{\centering\arraybackslash}m{4mm} 
    >{\centering\arraybackslash}m{4mm} 
    m{0mm}
    >{\centering\arraybackslash}m{6mm}%
    >{\centering\arraybackslash}m{6mm}%
}
\toprule
&&&&& \multicolumn{2}{c}{\textsc{Dir. Removed}} \\
\cmidrule{6-7}
& $\notin$ & N & R@10 && $\theta$ & $\Delta_{R@10}$ \\
\midrule
\textsc{Left} & 14 & 156 & 94.9\% && 0.0\% & -0.6\% \\[2mm]
\textsc{Front} & 13 & 486 & 81.1\% && 0.0\% & -0.6\% \\[2mm]
\textsc{Right} & 12 & 196 & 92.3\% && 0.0\% & -5.1\% \\[2mm]
\textsc{Back} & 9 & 564 & 81.6\% && 0.0\% & -2.3\% \\
\bottomrule
\end{tabular}
\end{table}

\subsection{Further details on Spatial Audio Caption Generation}
\label{sec:caption_gen_examples}

\begin{figure}[hp!]
    \centering
    \includegraphics[width=\textwidth, trim={30mm 143mm 53mm 146mm},clip]{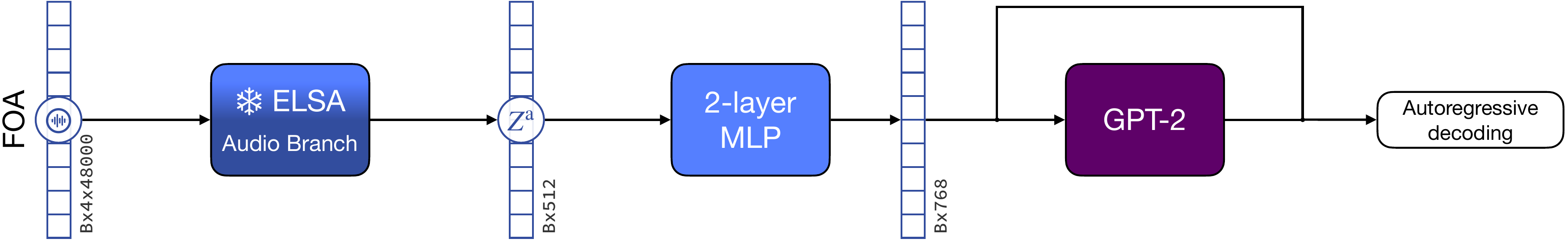}
    \caption{Architecture diagram for spatial audio caption generation.}
    \label{fig:caption_arch}
\end{figure}

\cref{sec:caption_gen} introduced a spatial audio caption generation system. In what follows, we illustrate some of the generations produced by the system as well as the corresponding ground-truth annotation. Below that, in \cref{fig:caption_arch}, we include an architecture diagram for the spatial audio caption generation system.

\vspace*{5mm}
\begin{captionenv} 
In a medium-sized room located at the far back, an electric motor is emitting a high-pitched whine, accompanied by a whirring noise. In the background, adult male voice can be heard speaking.
\mdfsubtitle{Ground-truth caption from test set}%
From deep within a medium-sized room, the noise of a robust industrial engine can be heard whirring loudly.
\end{captionenv}
\vspace*{5mm}

\begin{captionenv} 
The sound of water flowing and splashing is emanating from the front of a room.
\mdfsubtitle{Ground-truth caption from test set}%
The sound of gentle rowing and paddling in the water is emanating from the vicinity of a medium-sized room.
\end{captionenv}
\vspace*{5mm}

\begin{captionenv} 
The sound of cheering coming from a crowd is heard near the medium-sized room.
\mdfsubtitle{Ground-truth caption from test set}%
The sound of applause, indicating that people are praising the musicians after their performance, is emanating from the medium-sized room.
\end{captionenv}
\vspace*{5mm}

\newpage

\end{document}